\documentclass[aps,prc,twocolumn,superscriptaddress,showpacs,nofootinbib,floatfix]{revtex4}
\input epsf
\usepackage{graphicx}
\usepackage{dcolumn}
\usepackage{bm}
\newcommand{\pt} {\mbox{$p_T$\ }}
\newcommand{\mt} {\mbox{$m_T$\ }}

\def\Journal#1#2#3#4{{#1}{\bf #2}, #3 (#4)}

\def\NIMA{{Nucl. Instrum. Methods}~{\bf A}}
\def\NPA{{Nucl. Phys.}~{\bf A}}
\def\NPB{{Nucl. Phys.}~{\bf B}}
\def\PLB{{Phys. Lett.}~{\bf B}}
\def\PLC{Phys. Rep.\ }
\def\PRL{Phys. Rev. Lett.\ }
\def\PRD{{Phys. Rev.}~{\bf D}}
\def\PRC{{Phys. Rev.}~{\bf C}}
\def\ZPC{{Z. Phys.}~{\bf C}}

\def\ANP{Adv.~Nucl.~Phys}

\begin{document}
\title{Production of $\phi$ mesons at mid-rapidity in $\sqrt{s_{NN}}=200$~GeV Au+Au
collisions at RHIC}

\newcommand{\abilene}{Abilene Christian University, Abilene, TX 79699, USA}
\newcommand{\acadsin}{Institute of Physics, Academia Sinica, Taipei 11529, Taiwan}
\newcommand{\banaras}{Department of Physics, Banaras Hindu University, Varanasi 221005, India}
\newcommand{\barc}{Bhabha Atomic Research Centre, Bombay 400 085, India}
\newcommand{\bnl}{Brookhaven National Laboratory, Upton, NY 11973-5000, USA}
\newcommand{\caucr}{University of California - Riverside, Riverside, CA 92521, USA}
\newcommand{\ciae}{China Institute of Atomic Energy (CIAE), Beijing, People's Republic of China}
\newcommand{\cns}{Center for Nuclear Study, Graduate School of Science, University of Tokyo, 7-3-1 Hongo, Bunkyo, Tokyo 113-0033, Japan}
\newcommand{\columbia}{Columbia University, New York, NY 10027 and Nevis Laboratories, Irvington, NY 10533, USA}
\newcommand{\dapnia}{Dapnia, CEA Saclay, F-91191, Gif-sur-Yvette, France}
\newcommand{\debrecen}{Debrecen University, H-4010 Debrecen, Egyetem t{\'e}r 1, Hungary}
\newcommand{\fsu}{Florida State University, Tallahassee, FL 32306, USA}
\newcommand{\gsu}{Georgia State University, Atlanta, GA 30303, USA}
\newcommand{\hiroshima}{Hiroshima University, Kagamiyama, Higashi-Hiroshima 739-8526, Japan}
\newcommand{\ihepprot}{Institute for High Energy Physics (IHEP), Protvino, Russia}
\newcommand{\isu}{Iowa State University, Ames, IA 50011, USA}
\newcommand{\jinrdubna}{Joint Institute for Nuclear Research, 141980 Dubna, Moscow Region, Russia}
\newcommand{\kaeri}{KAERI, Cyclotron Application Laboratory, Seoul, South Korea}
\newcommand{\kangnung}{Kangnung National University, Kangnung 210-702, South Korea}
\newcommand{\kek}{KEK, High Energy Accelerator Research Organization, Tsukuba-shi, Ibaraki-ken 305-0801, Japan}
\newcommand{\kfki}{KFKI Research Institute for Particle and Nuclear Physics (RMKI), H-1525 Budapest 114, POBox 49, Hungary}
\newcommand{\korea}{Korea University, Seoul, 136-701, Korea}
\newcommand{\kurchatov}{Russian Research Center ``Kurchatov Institute", Moscow, Russia}
\newcommand{\kyoto}{Kyoto University, Kyoto 606, Japan}
\newcommand{\labllr}{Laboratoire Leprince-Ringuet, Ecole Polytechnique, CNRS-IN2P3, Route de Saclay, F-91128, Palaiseau, France}
\newcommand{\lawllnl}{Lawrence Livermore National Laboratory, Livermore, CA 94550, USA}
\newcommand{\losalamos}{Los Alamos National Laboratory, Los Alamos, NM 87545, USA}
\newcommand{\lpc}{LPC, Universit{\'e} Blaise Pascal, CNRS-IN2P3, Clermont-Fd, 63177 Aubiere Cedex, France}
\newcommand{\lund}{Department of Physics, Lund University, Box 118, SE-221 00 Lund, Sweden}
\newcommand{\muenster}{Institut f\"ur Kernphysik, University of Muenster, D-48149 Muenster, Germany}
\newcommand{\myongji}{Myongji University, Yongin, Kyonggido 449-728, Korea}
\newcommand{\nagasaki}{Nagasaki Institute of Applied Science, Nagasaki-shi, Nagasaki 851-0193, Japan}
\newcommand{\newmex}{University of New Mexico, Albuquerque, NM 87131, USA}
\newcommand{\nmsu}{New Mexico State University, Las Cruces, NM 88003, USA}
\newcommand{\ornl}{Oak Ridge National Laboratory, Oak Ridge, TN 37831, USA}
\newcommand{\orsay}{IPN-Orsay, Universite Paris Sud, CNRS-IN2P3, BP1, F-91406, Orsay, France}
\newcommand{\pnpi}{PNPI, Petersburg Nuclear Physics Institute, Gatchina, Russia}
\newcommand{\riken}{RIKEN (The Institute of Physical and Chemical Research), Wako, Saitama 351-0198, JAPAN}
\newcommand{\rkrbrc}{RIKEN BNL Research Center, Brookhaven National Laboratory, Upton, NY 11973-5000, USA}
\newcommand{\saispbstu}{St. Petersburg State Technical University, St. Petersburg, Russia}
\newcommand{\saopaulo}{Universidade de S{\~a}o Paulo, Instituto de F\'{\i}sica, Caixa Postal 66318, S{\~a}o Paulo CEP05315-970, Brazil}
\newcommand{\seoulnat}{System Electronics Laboratory, Seoul National University, Seoul, South Korea}
\newcommand{\stonybrkc}{Chemistry Department, Stony Brook University, SUNY, Stony Brook, NY 11794-3400, USA}
\newcommand{\stonycrkp}{Department of Physics and Astronomy, Stony Brook University, SUNY, Stony Brook, NY 11794, USA}
\newcommand{\subatech}{SUBATECH (Ecole des Mines de Nantes, CNRS-IN2P3, Universit{\'e} de Nantes) BP 20722 - 44307, Nantes, France}
\newcommand{\tenn}{University of Tennessee, Knoxville, TN 37996, USA}
\newcommand{\titech}{Department of Physics, Tokyo Institute of Technology, Tokyo, 152-8551, Japan}
\newcommand{\tsukuba}{Institute of Physics, University of Tsukuba, Tsukuba, Ibaraki 305, Japan}
\newcommand{\vandy}{Vanderbilt University, Nashville, TN 37235, USA}
\newcommand{\waseda}{Waseda University, Advanced Research Institute for Science and Engineering, 17 Kikui-cho, Shinjuku-ku, Tokyo 162-0044, Japan}
\newcommand{\weizmann}{Weizmann Institute, Rehovot 76100, Israel}
\newcommand{\yonsei}{Yonsei University, IPAP, Seoul 120-749, Korea}
\affiliation{\abilene}
\affiliation{\acadsin}
\affiliation{\banaras}
\affiliation{\barc}
\affiliation{\bnl}
\affiliation{\caucr}
\affiliation{\ciae}
\affiliation{\cns}
\affiliation{\columbia}
\affiliation{\dapnia}
\affiliation{\debrecen}
\affiliation{\fsu}
\affiliation{\gsu}
\affiliation{\hiroshima}
\affiliation{\ihepprot}
\affiliation{\isu}
\affiliation{\jinrdubna}
\affiliation{\kaeri}
\affiliation{\kangnung}
\affiliation{\kek}
\affiliation{\kfki}
\affiliation{\korea}
\affiliation{\kurchatov}
\affiliation{\kyoto}
\affiliation{\labllr}
\affiliation{\lawllnl}
\affiliation{\losalamos}
\affiliation{\lpc}
\affiliation{\lund}
\affiliation{\muenster}
\affiliation{\myongji}
\affiliation{\nagasaki}
\affiliation{\newmex}
\affiliation{\nmsu}
\affiliation{\ornl}
\affiliation{\orsay}
\affiliation{\pnpi}
\affiliation{\riken}
\affiliation{\rkrbrc}
\affiliation{\saispbstu}
\affiliation{\saopaulo}
\affiliation{\seoulnat}
\affiliation{\stonybrkc}
\affiliation{\stonycrkp}
\affiliation{\subatech}
\affiliation{\tenn}
\affiliation{\titech}
\affiliation{\tsukuba}
\affiliation{\vandy}
\affiliation{\waseda}
\affiliation{\weizmann}
\affiliation{\yonsei}
\author{S.S.~Adler}	\affiliation{\bnl}
\author{S.~Afanasiev}	\affiliation{\jinrdubna}
\author{C.~Aidala}	\affiliation{\bnl}
\author{N.N.~Ajitanand}	\affiliation{\stonybrkc}
\author{Y.~Akiba}	\affiliation{\kek} \affiliation{\riken}
\author{J.~Alexander}	\affiliation{\stonybrkc}
\author{R.~Amirikas}	\affiliation{\fsu}
\author{L.~Aphecetche}	\affiliation{\subatech}
\author{S.H.~Aronson}	\affiliation{\bnl}
\author{R.~Averbeck}	\affiliation{\stonycrkp}
% ...
\collaboration{PHENIX Collaboration} \noaffiliation

\date{\today}        

\begin{abstract}
We present the results of $\phi$ meson production in the $K^+K^-$ decay
channel from Au+Au collisions at $\sqrt{s_{NN}}=200$~GeV as measured at mid-rapidity
by the PHENIX detector at RHIC.
Precision resonance centroid and width values are extracted as a function of collision
centrality.  No significant variation from the PDG accepted values is observed,
contrary to some model predictions.  The $\phi$ transverse
mass spectra are fitted with a linear exponential function for which the derived inverse
slope parameter is seen to be constant as a function of centrality.  However,
when these data are fitted
by a hydrodynamic model the result is that the centrality-dependent
freeze--out temperature and the expansion velocity
values are consistent with the values previously derived from fitting identified charged hadron data.
As a function of transverse momentum the collisions scaled
peripheral--to--central yield ratio $R_{CP}$ for the $\phi$ is comparable to
that of pions rather
than that of protons. This result lends support to theoretical models which
distinguish between baryons and mesons instead of particle mass for explaining
the anomalous (anti)proton yield.
\end{abstract}
\pacs{25.75.Dw}
\maketitle

\parskip 3pt

\section{INTRODUCTION}
\label{sec:intro}
Relativistic heavy-ion experiments have a goal of producing
matter at extreme temperatures and energy densities such that
conditions are favorable for the transition to a deconfined
state of quarks and gluons, the Quark Gluon Plasma (QGP).
Theoretical calculations predict that the temperatures
and energy densities which can be reached at the Brookhaven
National Laboratory's Relativistic Heavy Ion Collider (RHIC)
will exceed those needed for the formation of the
QGP~\cite{LaermannAndPhilipsen03, Bjorken83, Adcox01}.

The production and decay of the $\phi$ meson have long been recognized
as an important probe for the state of matter produced in relativistic
heavy ion (RHI) collisions~\cite{RafelskiAndMueller82, Shor85, Koch86,
Singh86, Singh87, Singh93, Singh97, Bass99,
BlaizotAndGalain91, Pal2002,
BiAndRafelski91,
AsakawaAndKo94, Song96, AsakawaAndKo94a,
KoAndSiebert94, Haglin95, 
SmithAndHaglin98, 
OsetAndRamos01}.  In
$pp$ collisions the creation of the $\phi$ is
suppressed according to the Okubo-Zweig-Iizuka rule~\cite{Okubo63}.
Hence, if there is an enhancement of the $\phi$ yield
in RHI collisions relative to $pp$ collisions, this could be evidence of
non-conventional production mechanisms such as strange quark
coalescence via the formation and subsequent hadronization
of the QGP.  The fact that the $\phi$ yield is undistorted by feed-down 
from higher mass resonances makes it an attractive probe in this respect.

The decay modes of the $\phi$, specifically the dilepton channels
($e^+e^-$ or $\mu^+\mu^-$) and the $K^+K^-$ channel, will probe
the final state differently should the decay take place in
the presence of the QGP-mixed or the completely hadronized phase.  The dileptons
will have insignificant interactions with the medium, while
the kaons can scatter until freeze-out.  The lifetime of the
$\phi$ in vacuum is large ($\approx$~45~fm/$c$) compared to
say a 10~fm sized interaction region.  However, several
theoretical calculations~\cite{Pal2002, OsetAndRamos01, CabreraAndVacas03}
predict that the $\phi$ mass and width could be significantly
modified in either the hot or the cold nuclear medium.
These medium induced effects could be manifested through measured shifts in the
mass centroid of the resonance or changes in the resonance width.  Also
predicted are changes in the relative branching ratio between kaon and
lepton pairs, with respect to the Particle Data Group (PDG~\cite{PDG04}). 

The production mechanism of strangeness in heavy ion collisions can be investigated 
through the measurement of the particle yields. In this paper, we study system size dependence by 
analyzing centrality selected data. A comparison between different systems can be made
by normalizing to the number of participant pairs. The expectation is that for production 
dominated by soft processes, the yields scale as the number of participants. We
compare the centrality dependence of 
strange and non-strange particle yields in order to reveal the possible flavor
dependence.  

An additional important question is whether the $\phi$ mesons participate 
in radial flow together with the other hadrons, or if they freeze-out earlier,
as might be true if the small vacuum cross sections of the $\phi$ with
hadrons persist in the fireball. Previous measurements have yielded 
contradictory results~\cite{NA50,NA49}.
One of the important advantages of RHIC experiments is the capability
to examine the momentum spectrum as a function of centrality for a
variety of hadrons which should yield important additional information
on the radial flow issue. A spectral shape analysis including a simultaneous treatment 
of the $\phi$ and the more abundant hadrons ($\pi,K,p$) will be presented here.     

At high $p_{T}$, hadrons are primarily produced from the fragmentation of hard-scattered partons. One of the most
exciting results from RHIC was the discovery of hadron suppression in central
Au+Au collisions~\cite{ppg003,ppg013} where this suppression is absent
in d+Au collisions~\cite{ppg014}.
Surprisingly, it was also discovered that proton and anti-proton production at
intermediate $p_{T}$ (1.5--4.5~GeV/$c$) scales 
with the number of binary nucleon-nucleon collisions ($N_{coll}$) as
would be expected for hard-scattering in the absence of any 
nuclear modification~\cite{protonscaling}.  In fact,
the intermediate $p_{T}$ anti-proton to pion ratios
were found to exceed by a 
factor of~3 the values expected from parton
fragmentation~\cite{protonscaling,ppg006}. These experimental results 
lead to the conclusion that protons and pions have different
production mechanisms at intermediate $p_{T}$~\cite{protonscaling}. One possible
explanation invokes parton recombination from
the QGP~\cite{Fries03a, Hwa:2004ng, Greco:2003mm, Greco:2003xt}. A measurement 
of the nuclear modification factor for the $\phi$ meson, which has a mass
comparable to the proton but carries only two 
quarks, is crucial for understanding the hadron production at intermediate
$p_{T}$.
In this work we have measured the nuclear 
modification factor through the ratio $R_{CP}$ of central to peripheral yields
scaled by their respective $N_{coll}$~value.        

To put our results into perspective, we begin by describing the currently available
$\phi$ data obtained in RHI collisions. The production of $\phi$~mesons has been studied
systematically at ever increasing $\sqrt{s}$ from the AGS to RHIC.
The E802 collaboration made the first observation of the $\phi$
in fixed target central collisions of 14.6$A$~GeV~Si+Au
($\sqrt{s_{NN}}=$~5.39~GeV) via
the $K^+K^-$ channel~\cite{Akiba96}.
They obtained a ratio $N_\phi/N_{K^-} = 11.6\%$, roughly consistent
with the ratio obtained in $pp$ data over a wide range of
$\sqrt{s}$~\cite{Fesefeldt79}.  The analysis of the E802 rapidity
distributions indicated that the $\phi$ production scaled with
the product of the $K^+$ and $K^-$ separate yields, and that
there was either significant rescattering of the $\phi$
after production or the production itself came after
rescattering of the colliding participants.

Also at the AGS the E917 experiment
has reported another $\phi$ measurement
with 11.7A~GeV Au+Au ($\sqrt{s_{NN}}=$~4.87~GeV)  in the rapidity range
$1.2 < y < 1.6$ in five centrality bins~\cite{Back03}
The observed
yield of the $\phi$ increased towards more central collisions
with a distinctly faster than linear dependence on the number of
participants.  The yield increase of the $\phi$ in central
collisions was stronger than that of the $\pi$ since the
$N_{\phi}/N_{\pi}$ ratio increased in central collisions.  However,
the $N_{\phi}/N_{K^+}$ and $N_{\phi}/N_{K^-}$ ratios were
constant as a function of centrality.

At the SPS the NA49 experiment has measured $\phi$ production
in $pp$, $p$+Pb, and Pb+Pb collisions with E$_{\hbox{beam}}$=158$A$~GeV
($\sqrt{s_{NN}}=$~17.5~GeV)
in the rapidity range $3.0 < y < 3.8$~\cite{Afanasiev00}.
Relative to the $pp$
yields, these data showed that the ratio of the $\phi$ yield
to the $\pi$ yield in 
central Pb+Pb collisions was enhanced by
a factor $3.0\pm 0.7$.  Another SPS collaboration, NA38/NA50, has
measured the $\phi$ in the $\mu^+\mu^-$ channel~\cite{Abreu96,
Abreu01}, for which the extracted effective temperature and $dN/dy$
differ from those obtained in
the $K^+K^-$ channel in the same systems. The yield
difference between the two SPS experiments has been
calculated to be factors of 2--4~\cite{Rohrich01}, with the NA38/NA50
result being higher.

The first measurement of the $\phi$ meson at RHIC was reported
by the STAR collaboration~\cite{Adler02},
in the collisions of Au+Au at
$\sqrt{s_{NN}}$=130 GeV at three centralities, 0--11\%, 11--26\%,
and 26--85\% in the rapidity range $-0.5 < y < +0.5$.
The extracted temperature $T$ and the ratio $N_\phi/N_h$
did not vary with centrality.  

One may summarize the current state of knowledge of $\phi$ production
in heavy ion collisions by stating that the topic remains highly unexplored
territory.
The heavy-ion measurements do indicate
that the observed $\phi$ yield is not a simple linear superposition of nucleon-nucleon
collisions.  Rather the data imply the influence of some collective
effects.  Whether those effects are induced by cold and/or hot nuclear matter,
there do not yet exist definitive measurements. 
Moreover, there has not been so far precise enough
heavy-ion data which can address the question of the change in the $\phi$
mass or its width in the cold nuclear or the hot QGP medium.
And except for one experiment measuring the dimuon channel, there is a
scarcity of useful quantitative
information in heavy ion collisions concerning the $\phi$ decay
into dileptons.

In this paper we report on a measurement of the $\phi$ yield
at mid-rapidity in collisions of Au+Au beams
from RHIC at $\sqrt{s_{NN}}=200$~GeV as measured in the $K^+K^-$ channel by the
PHENIX detector.
The paper is organized as follows.  In Section II we give a short review
of the PHENIX detector configuration.  In Section III we describe the
data analysis procedure.  Section IV presents and discusses the
results.  The first precision measurements of $\phi$ mass and width
values obtained in relativistic heavy ion collisions as a
function of centrality are given in Section IV--A.
Absolute yields as a function of \pt for three centrality
bins are shown in Section IV--B.
The centrality 
dependence of yields and ratios are studied in Section~IV--D. In Section IV--D
the spectra shapes are interpreted in the framework of a 
hydrodynamical model and the freeze-out conditions are extracted.
Finally, the nuclear modification factor $R_{CP}$ 
for the $\phi$ is obtained and compared to those of pions and protons in Section~IV--E.

\section{PHENIX Detector}
\label{sec:phenix}
\begin{figure}
{\includegraphics[width=1.0\linewidth]{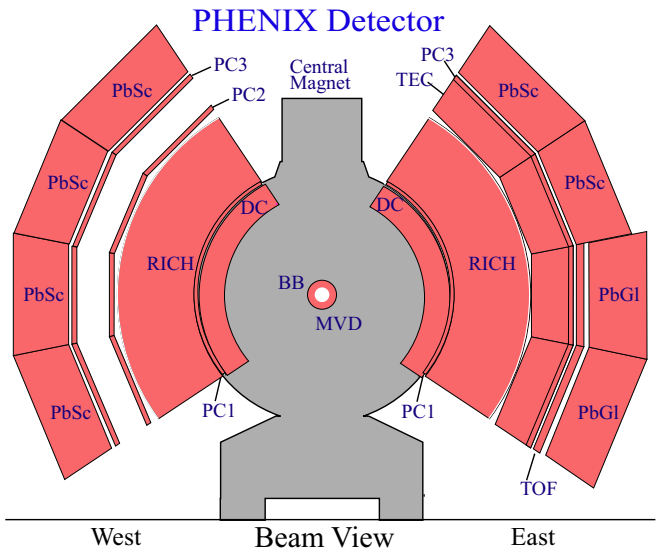}}
\caption{\label{fg:Phenix_2002_beam} Beam's eye view of the PHENIX central arm
detector subsystems.}
\end{figure}

The PHENIX detector consists of two spectrometer arms at
near zero rapidity, two forward rapidity muon spectrometers, and
three global event characterization detectors. The central arm
spectrometers, shown schematically in
Fig.~\ref{fg:Phenix_2002_beam}, are located East and West of the beam line with
$\pi/2$ radian azimuthal coverage each.  These spectrometers are designed to detect
photons, electrons, and charged hadrons.  The $\phi$ data for this paper
were obtained with the central arm
detector subsystems which provide high resolution particle identification
and momentum reconstruction.
A complete description
of the PHENIX apparatus has been published
elsewhere~\cite{PHENIX_overview, PHENIX_PID,PHENIX_EMC,PHENIX_inner,PHENIX_ZDC,PHENIX_tracking}.
We present a brief review of the relevant detector subsystems in the following
sections.
\subsection{Global Detectors}
The global detectors furnish the start time signals, collision vertex
measurements, and interaction centrality. 
The centrality  for events in the Au+Au collisions is determined~\cite{ppg026}
by combining the data from two subsystems: the Zero Degree Calorimeters
(ZDC)~\cite{PHENIX_ZDC} and the Beam-Beam Counters (BBC)~\cite{PHENIX_inner}.
The ZDC are hadronic calorimeters located 18~m downstream and
upstream of the interaction point along the beam line.
These calorimeters detect the energy carried by
spectator neutrons.  The BBC are $\check{\rm C}$erenkov telescopes placed
$\pm 1.44$~m the center of the beam collision region in the
pseudo-rapidity region 3.0~$<$~$|\eta|$~$<$~3.9.  The correlation
between the ZDC energy sum and the charge sum recorded in the BBC
determines the centrality of the collision event.  The BBC data also determine
the longitudinal collision coordinate ($z_{\hbox{vertex}}$) and the start time
for the time-of-flight measurements.

\subsection{Central Arm Detectors}
The central arm spectrometers~\cite{PHENIX_tracking} in PHENIX
provide charged particle tracking
and particle identification.  This analysis was done with the East arm
spectrometer.  The $\phi\rightarrow K^+K^-$ decay kinematics are such that
that the PHENIX detector has negligible acceptance for the very low $p_T$
$\phi$ particles which would decay into East-West kaon pairs. 
The data included information from the drift chamber (DC), the pad chambers
(PC1 and PC3), the high resolution time-of-flight wall
(TOF), and the electromagnetic calorimeter (EMCal) lead-scintillator
detectors (PbSc), as depicted in Fig.~\ref{fg:Phenix_2002_beam}.

Pattern recognition and tracking of the charged particles are accomplished 
using the DC information by a combinatorial Hough transform 
method~\cite{hough}. The DC subsystem is located at an average
radial distance of 2.2~m from the beam line. It is a projective
tracking detector providing high precision measurements in the
azimuthal $XY$ plane, which are combined with the $XYZ$ single
spatial point measurement from the PC1 located at $2.45$~m.
These data, together with the BBC $z_{\hbox{vertex}}$ information, are
sufficient to determine the track's initial momentum vector
whose magnitude is obtained with a resolution
$\delta p/p \simeq 0.7\% \oplus 1.0\%\times p$ (GeV/$c$).
The first term in this expression is due to the multiple scattering before the DC and the 
second term is the angular resolution of the DC.  Based on identified
mass measurements obtained with the TOF subsystem, the absolute
momentum scale is known to $\pm 0.7\%$.

Tracks obtained from the DC/PC1 detectors are projected to the
PC3, TOF, and PbSc detectors where associations can be made.
The high resolution TOF subsystem provides one set of
mass measurements while the PbSc detectors provide a geometrically
independent set of mass measurements.  The TOF wall is positioned
5.06~m from the beam line and consists of 960 scintillator slats oriented along 
the azimuthal direction. It is designed to cover $|\eta|< 0.35$ and 
$\Delta\phi=\pi/4$ in azimuthal angle. 

The PbSc detector, covering half of the East arm and entire West arm,
can also be used for hadron timing measurements. The present analysis
uses the PbSc modules in the East Arm which are located at 5.1~m in radial
distance from the beam line and cover a $\Delta\phi = \pi/4$ azimuthal range.
This detector is constructed
as separate towers of dimension 5.25~x~5.25~x~37~cm$^3$, in an alternating lead-scintillator
sandwich type structure (``shish-kebob''), approximately
18~radiation lengths in depth. As illustrated in Fig.~{\ref{fg:Phenix_2002_beam}}
the TOF and the PbSc sectors are completely non-overlapping.  

\section{DATA ANALYSIS}
\label{sec:analysis}
In this section, we describe the event and track selection, particle
identification, the details of $K^{+}K^{-}$ pair reconstruction, and the corrections
for geometrical acceptance, particle decay in flight, multiple scattering,
and detector occupancy factors, all of which couple into deriving the $\phi$ meson spectra.
\subsection{Event Selection}
\label{sec:event_selection}
The events selected for this analysis were
based on the PHENIX minimum-bias trigger provided 
by the beam-beam counters (BBC) and zero-degree calorimeters (ZDC). As noted
previously, the
centrality of each Au + Au collision event 
was determined by correlating the BBC charge sum and the ZDC total
energy~\cite{ppg026}. The PHENIX minimum-bias data sample 
included $92.2^{+2.5}_{-3.0}$~\% of the 6.9 barn Au + Au total inelastic cross
section~\cite{ppg014}.  This analysis used 20~million minimum-bias 
events with a vertex position within $|z_{\hbox{vertex}}| < 30$ cm. 

To study the centrality dependent physics, we divided these events into
different centrality bins.  
For the $\phi$ meson line shape analysis, we used five
centrality bins: 0--10\%, 10--20\%, 20--40\%, 40--60\% and 60--92\%. 
The transverse mass ($m_T$)  spectra were reconstructed for three 
centrality bins: 0--10\%, 10--40\% and 40--92\%. These bin divisions were 
chosen to have approximately equal statistical significance for their respective
data points. 
The centrality of collisions is additionally characterized
by the average number of participants ($<N_{part}>$) and the average number of
binary nucleon--nucleon collisions ($<N_{coll}>$).
These two global quantities, shown in Table~\ref{ncoll} as a function of
centrality, are derived from a 
Glauber model calculation~\cite{ppg014}. 

\begin{table}
\caption{\label{ncoll} Average number of participants and collisions in Au + Au
reaction at RHIC for different centralities determined from a Glauber
model\cite{ppg014}. 
The error associated with each number is the systematic error.}
\begin{center}
\begin{ruledtabular}
\begin{tabular}{ccc}
~~~Centrality~~~ &~~~~~ $<N_{part}>$~~~~~ & ~~~~~$<N_{coll}>$~~~~~ \\
(\%) & &\\\hline
0 - 10 & 325.2 $\pm$ 3.3  & 955.4 $\pm$ 93.6 \\
10 - 20 & 234.6 $\pm$ 4.7 & 602.6 $\pm$ 59.3 \\
20 - 40 & 140.4 $\pm$ 4.9 & 296.8 $\pm$ 31.1 \\
10 - 40 & 171.8 $\pm$ 4.8 & 398.7 $\pm$ 40.5 \\
40 - 60 & 59.9 $\pm$ 3.5 & 90.6 $\pm$ 11.8 \\
60 - 92 & 14.5 $\pm$ 2.5 & 14.5 $\pm$ 4.0 \\
40 - 92 & 32.0 $\pm$ 2.9 & 45.2 $\pm$ 7.3 \\
Min. bias & 109.1 $\pm$ 4.1 & 257.8 $\pm$ 25.4 \\
\end{tabular}
\end{ruledtabular}
\end{center}
\end{table}
\subsection{Track Selection}
\label{sec:track}
Only tracks with valid information from the DC and the PC1 were used for the 
analysis. These tracks were then confirmed by matching the projected and 
associated hit information at the
TOF wall for the TOF analysis, or at the PC3 and the EMCal for the PbSc
analysis.
The differences between the actual azimuthal and longitudinal hit
coordinates compared to the projected hit coordinates were determined.
These tracking coordinate residuals were converted to standard deviation residuals
by a momentum dependent function which computed the expected residual
coordinate value.
On this basis, a~3$\sigma$~track matching cut was used to accept track associations.
Lastly, for the TOF wall, an energy loss cut is applied on the analog
signal height from the scintillator slat. This cut has been described in a previous
publication~\cite{ppg026}.
\subsection{Particle Identification}
\label{sec:pid}
\begin{figure}[t]
\includegraphics[width=1.0\linewidth]{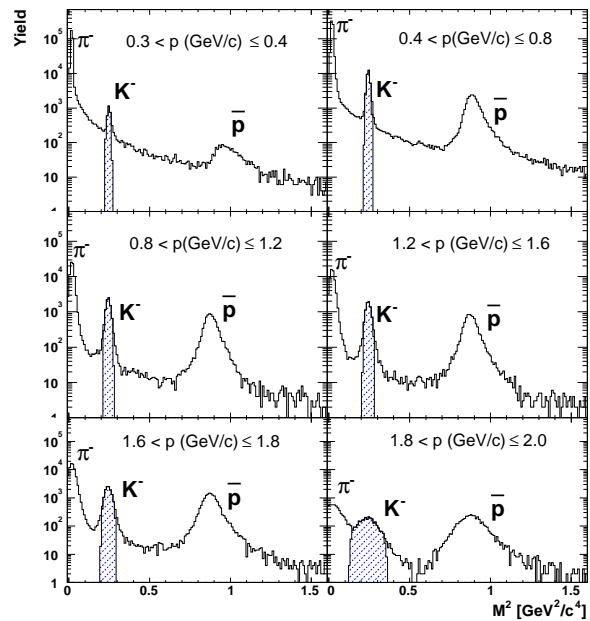}
\caption{\label{tofpid} Mass-squared distribution of all selected tracks passing through the TOF
for six selected momentum bins.
The identified kaons within $2\sigma$ mass--squared boundaries
at the different momentum bins are shown by
the shaded region on each plot.}
\end{figure}

As mentioned earlier, the PHENIX central arm spectrometer utilizes the high resolution
TOF wall and PbSc modules for hadron mass identification. 
The kaons in the TOF wall were identified via reconstructed momentum
combined with a time of flight measurement.
The timing resolution of this subsystem is $\sigma \simeq 115$~ps.
A momentum range of 0.3--2.0~GeV/$c$ was selected in order to compute the mass
distributions~\footnote{The mass-squared for each
track is defined as, $M^{2}~=~p^{2} (~t^{2}c^{2}/L^{2}~-~1)$, where $p$ is the momentum, 
$t$ is the time of flight, $c$ is the speed of light and $L$ is the length of the path traversed by
the track from vertex to the detector.}. Fig.~\ref{tofpid}  shows the
mass-squared distribution of all tracks passing through the TOF module for six different 
momentum bins. The kaons were
identified by applying a $2\sigma$ mass--squared cut, which is shown by the shaded region in each plot.


\begin{figure}[t]
\includegraphics[width=1.0\linewidth]{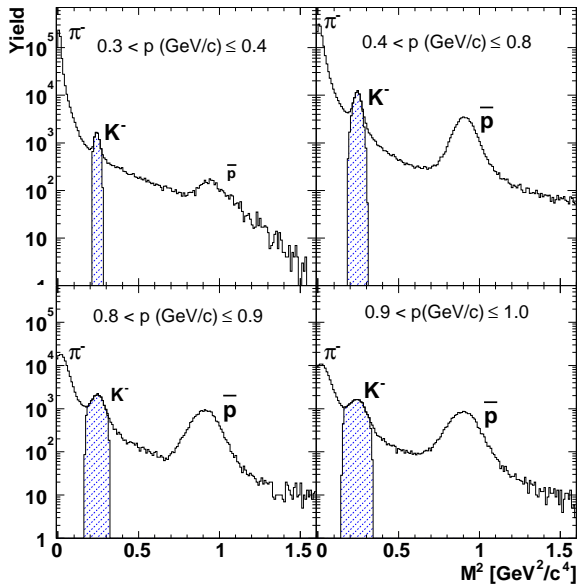}
\caption{\label{emcpid} Mass-squared distribution of all selected tracks
passing through the PbSc 
for four momentum bins. The identified kaons within $2\sigma$ mass--squared boundaries 
at the different momentum bins are shown by
the shaded region on each plot.}
\end{figure}

The particle identification with the electromagnetic calorimeter modules
is sensitive to the fact that
the electromagnetic and hadronic interactions 
produce quite different patterns of energy sharing between calorimeter
towers~\cite{emcalnim}.  As a consequence, the hadron timing properties
of the PbSc depend on the energy deposited on the central tower
of the cluster, particle momentum, particle type, charge, angle of
incident of the track, {\it etc}. The PbSc hadronic timing response was corrected for
these effects and we obtained an overall timing resolution of $\sigma \simeq450$~ps which 
is sufficient to enable a clear $\pi/K$ separation within $0.3<p\hbox{\ (GeV/$c$)\ }<1.0$
using a $2\sigma$ mass--squared selection criterion.
In Fig.~\ref{emcpid}  the mass-squared distributions are plotted for four
different momentum slices
for all tracks passing through PC3 and PbSc. 
The identified kaons are also shown in the figure by the $2\sigma$~width
shaded histograms superimposed on
the $M^{2}$~distributions for all tracks in different momentum bins. 

\subsection{$K^{+}K^{-}$ Invariant Mass spectra and $\phi$ Signal Extraction}
All identified $K^{+}$ and $K^{-}$ tracks in a given event were 
combined to form the invariant pair mass distributions. Three different pair
combinations were used. These are
\begin{itemize}
\item [a)] Both $K^{+}$ and $K^{-}$ identified by  the TOF detector (TOF--TOF combination),
\item [b)] $K^{+}$ identified with TOF and $K^{-}$ identified with PbSc (TOF--PbSc 
combination), and
\item [c)] Both $K^{+}$ and $K^{-}$ identified with PbSc detector (PbSc--PbSc
combination).
\end{itemize}
We did not use the $K^{+}$ from PbSc and $K^{-}$ from TOF in the
b)~combination. This is due to the fact that the PHENIX central arm geometry, in
the presence of a 0.8~T-m magnetic field,
does not have any acceptance for such pairs below an invariant
mass~of~1.06~GeV/$c^{2}$ in the TOF--PbSc combination.

A large combinatorial background is inherent to the $K^{+}K^{-}$ pair
invariant mass distribution. The combinatorial background was estimated by an
event mixing method in which all $K^{+}$ tracks from one event were
combined with $K^{-}$ tracks of ten other events within the same centrality
and vertex class.
The mixed event technique reproduces the shape of the unlike sign combinatorial
background. Finally, the size of the combinatorial
background is obtained by normalizing the mixed event invariant mass
spectra to 2$\sqrt{N_{++}N_{--}}$ where $N_{++}$ and $N_{--}$ represent
the measured yields in $K^{+}K^{+}$ and $K^{-}K^{-}$ mass distributions
respectively. This normalization is derived analytically
starting from the assumption
that the number of $K^{\pm}$ tracks per event
follows a Poisson distribution. A complete derivation
of this is given in Appendix~A.

\begin{figure*}[t]
\includegraphics[width=1.0\linewidth]{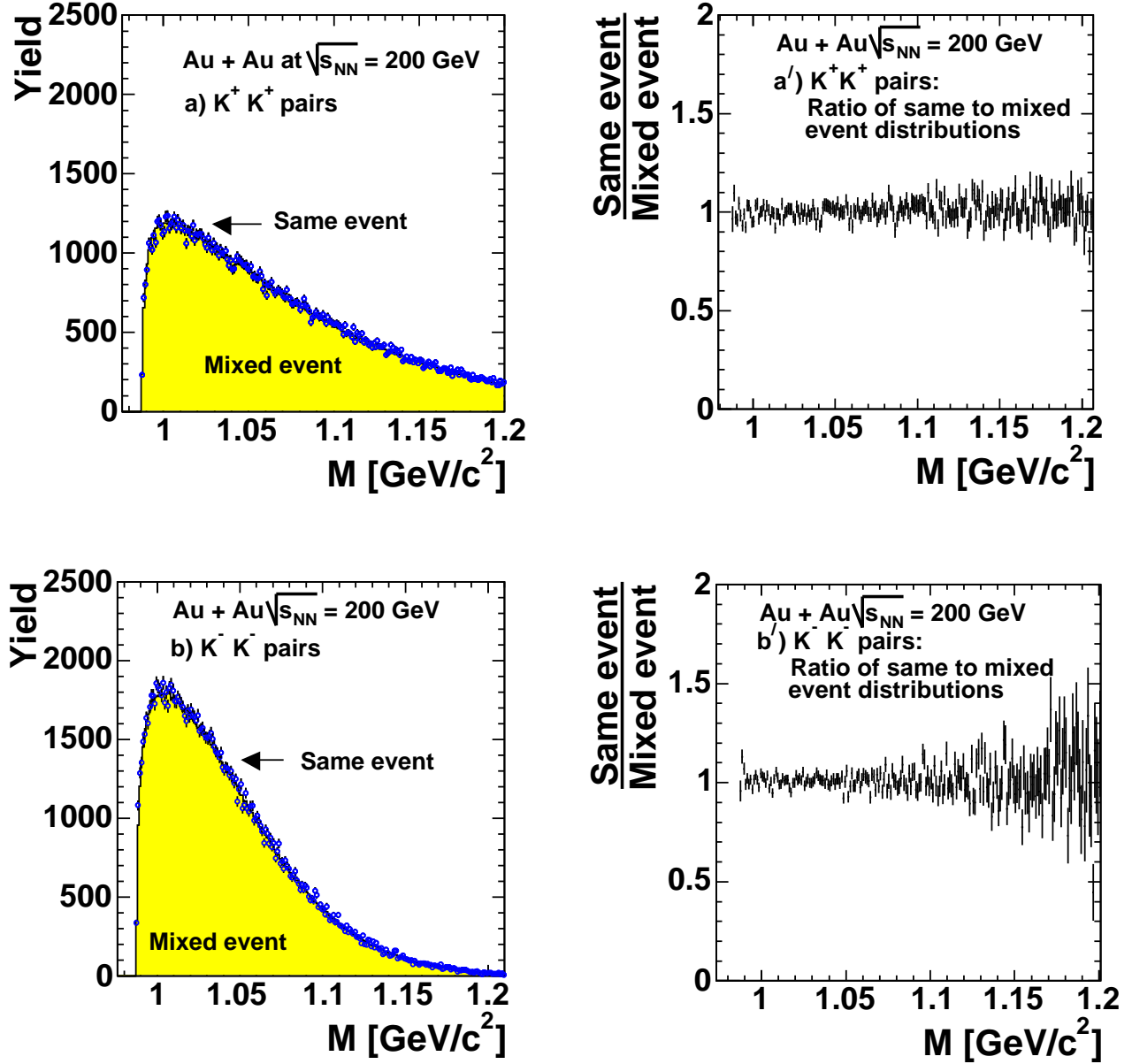}
\caption {\label{likesign} $K^{+}K^{+}$ [(a) and (a$^\prime$)] and $K^{-}K^{-}$
[(b) and (b$^\prime$)] invariant
mass spectra and ratio of real to mixed event spectra, respectively, for the
TOF--PbSc combination.}
\end{figure*}

The ability of this event mixing method to
reproduce correctly the shape of the combinatorial background distribution was
confirmed by constructing, in a similar way, the mixed--event like-sign
spectrum and comparing it to the same--event like-sign pair distribution.
The assumption is that the like-sign pair distributions are purely
combinatoric. For the three detector combinations
TOF--TOF, TOF--PbSc and PbSc--PbSc, the
ratio of the measured and combinatorial like-sign invariant mass distributions
were found to be consistent with 1.0 as a function of the pair mass within 
statistical errors for all centrality bins.
As an example, in Fig.~\ref{likesign} we plot the measured and 
combinatorial ``+\,+'' and ``-\,-'' invariant mass distributions and their ratios for
the TOF--PbSc combination as a function 
of the invariant mass of $K^{+}K^{+}$ and $K^{-}K^{-}$ pairs for
minimum-bias events. As can be seen from the figures, 
these ratios are equal to the expected value of~1.0 within the statistical
fluctuations.

The systematic uncertainty associated with this normalization
procedure was estimated to vary between 0.5--2\% for the different centralities
in the different detector combinations. When we added all data together to
derive the $\phi$ spectrum, the range of the systematic
uncertainty reduced to 0.7--1.0\%.

\begin{figure}[t]
\includegraphics[width=1.0\linewidth]{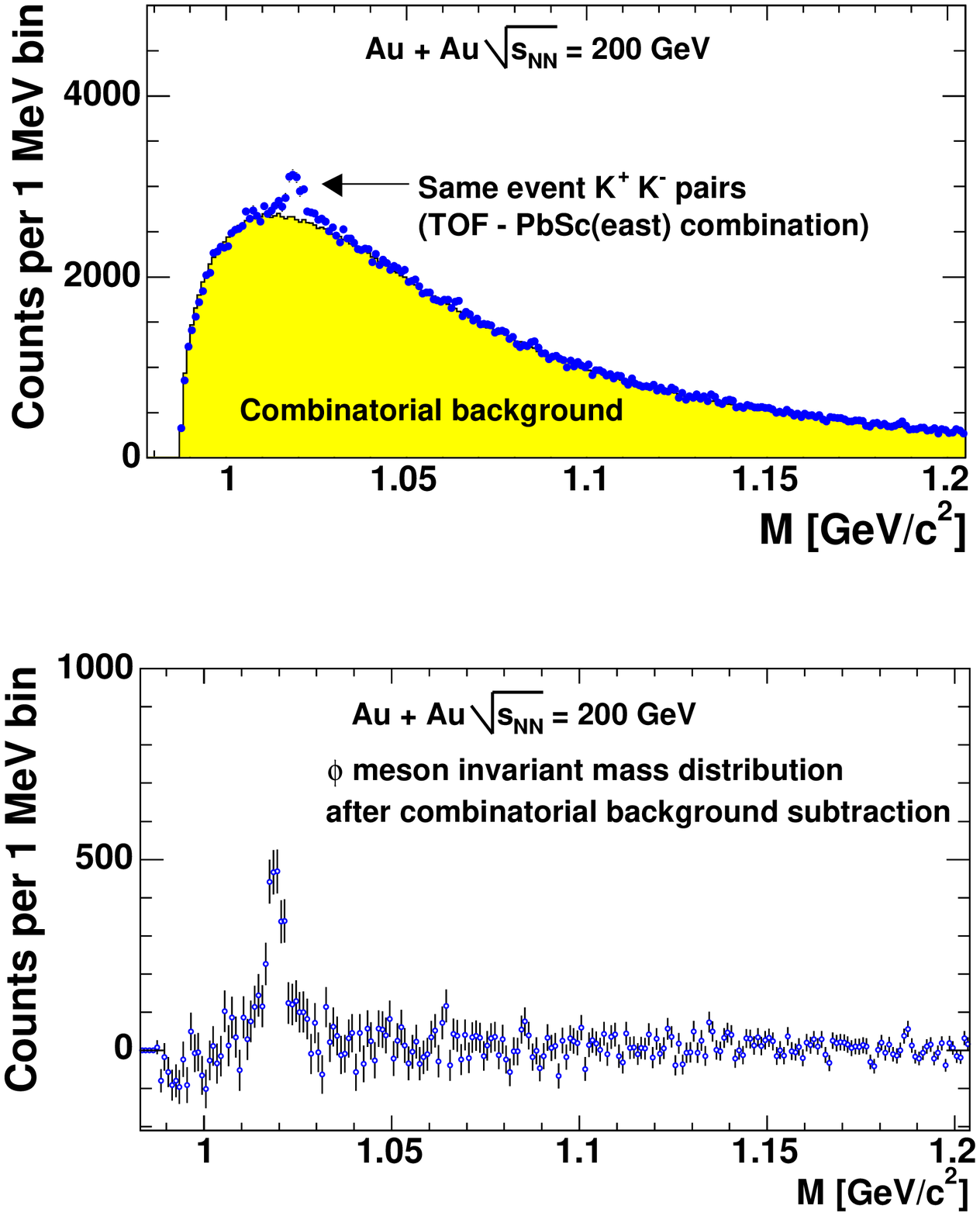}
\caption {\label{inv-example} $K^{+}K^{-}$ invariant mass spectra for the measured and mixed
events (top half) in the TOF-PbSc combination,
and the subtracted mass spectrum showing the $\phi$ meson peak
clearly above the background (bottom half).}
\end{figure}

Finally, the $\phi$ meson signal was obtained by subtracting the combinatorial
background from the measured unlike-sign invariant mass spectrum. An example of the
$K^{+}K^{-}$ invariant mass spectrum for the TOF--PbSc combination is shown in 
Fig.~\ref{inv-example} where we plotted the measured and scaled mixed event 
invariant mass distributions for  
minimum-bias events. The lower panel of
the figure shows the subtracted mass spectrum. The corrected yield of
the $\phi$ mesons from the experimental data is then determined by
integrating the subtracted invariant mass spectrum within a mass window
of $\pm$~5~MeV/$c^2$ about the measured $\phi$ mass centroid.  This narrow mass window
was used as it provided a better signal-to-background ratio compared with
a wider window.  Since we will show that there is no significant centrality
dependence of the intrinsic width, then the extracted yields as a function of
centrality are not being biased by the use of a constant integration window.
The systematic effect of the mass integration window itself on the corrected
yield was studied by varying the stated integration limit and found to contribute 2.6--3.2\%, depending
on centrality, to the total systematic uncertainty in the integrated yield.

\subsection{Acceptance, Decay and Multiple Scattering Corrections}
\label{sec:MC}
The $\phi$ meson yields were corrected for the geometrical acceptance of the
detectors, in-flight kaon decay, multiple scattering effects, and nuclear 
interactions with materials in the detector using the PISA software
package which is a GEANT-based~\cite{geant}
Monte Carlo detector simulation of the
PHENIX detector. The simulation was carried out by
generating 34~million single $\phi$ mesons in a
$\pm 0.6$~rapidity interval with an exponential transverse momentum 
distribution
\begin{equation}
dN/dp_T = p_T exp (-m_T/(t_{fo} + \beta^2 M_{\phi}))
\end{equation}
with $t_{fo} = 157$~MeV and $\beta = 0.4$, {\it i.e.}~an effective
slope of $T = 320$~MeV
in the range $0 <  p_{T}< 10$~GeV/$c$.
The generated $\phi$ 
mesons were then propagated through the simulation package.
In this simulation, the BBC, DC, PC, TOF and PbSc detector responses were
tuned to match the real data by including their dead areas and by matching 
their track associations and mass--squared resolutions. That is, the track association and
mass--squared cut boundaries in the Monte Carlo analysis were parameterized to match the real data.
The $K^{+}K^{-}$ pair acceptance efficiency as a function of
transverse mass was calculated as

\begin{equation}
\epsilon (m_{T}) = \frac{N^{reconstructed}_{\phi}(m_{T})}{N^{generated}_{\phi}(m_{T})}
\end{equation}

\begin{figure}[t]
\includegraphics[width=1.0\linewidth]{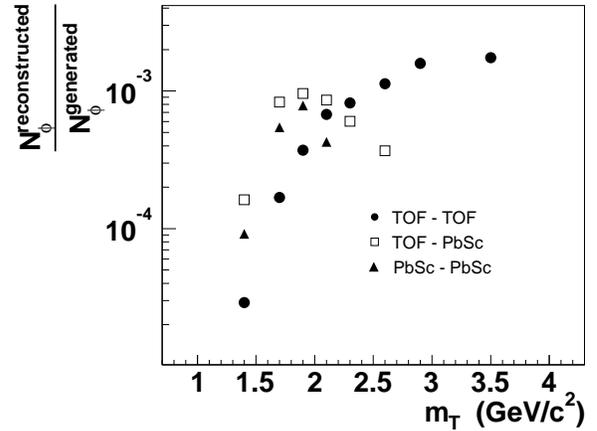}
\caption {\label{eff} Detector acceptance efficiency vs transverse mass of
the simulated $K^{+}K^{-}$ pairs for the TOF--TOF, TOF--PbSc
and PbSc--PbSc combinations. The statistical errors are smaller than
the size of the data symbols.}
\end{figure}

The calculated acceptance efficiencies for the TOF--TOF, TOF--PbSc and PbSc--PbSc
combinations are shown
in Fig.~\ref{eff} as a  function
of $m_{T}$. The points in the figure are located at the center of the bins. 
In the actual $m_{T}$~spectra, the proper bin centroids were used. The
figure shows that the TOF detector (closed circles) has low acceptance for the low 
momentum kaon pairs due to their large opening angles.  On the other hand,
the TOF covers the largest
$m_{T}$~range for the $\phi$ particles. As a result, the TOF efficiency
function increases 
towards higher transverse mass. In contrast, the TOF--PbSc (open squares) and
PbSc--PbSc (closed triangles)
combinations offer better low pair momentum acceptance than the TOF.
However, the high momentum kaon identification limit in the PbSc leads to the
efficiency function decreasing at the highest transverse mass values.
The systematic error associated with the acceptance correction factor
originates from
\begin{itemize}
\item [i)] tuning of detector alignments and mass--squared parameters in the Monte Carlo with reference to
the real data ($\sim$ 3\%), and
\item [ii)] systematics in the fiducial geometries in the
data and the Monte Carlo ($\sim$ 12\%).
\end{itemize}

 \subsection{Detector Occupancy Correction}
\label{sec:emb}
\begin{figure}[t]
\includegraphics[width=1.0\linewidth]{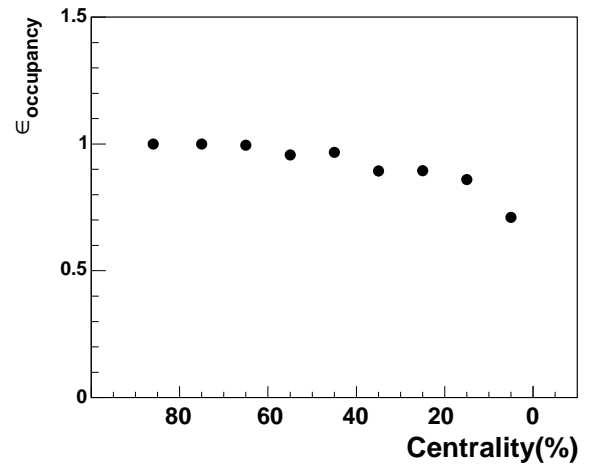}
\caption {\label{emb} Multiplicity (occupancy) dependent efficiency correction for
detecting the $\phi$~meson in the PHENIX detector as
a function of the collision centrality. The most central collisions are to the
right, the most peripheral collisions are to the left.  The statistical errors are
less than the size of the data points.}
\end{figure}

The high multiplicity environment in the heavy ion collisions produces
multiple hits in a detector cell such as in the slats of the TOF or
in the towers of the PbSc. These occupancy effects reduce the track reconstruction efficiency in
central collisions compared to that in peripheral collisions,
and these occupancy dependent effects need to accounted
for in calculating the invariant yields. The
multiplicity dependent efficiency ($\epsilon_{occupancy}$) factors were calculated by 
embedding simulated $K^{+}K^{-}$ pairs into real data events. This study was 
done for different  centrality bins from 0 to 92\% in steps of 
10\%. We calculated the multiplicity dependent efficiencies for TOF--TOF, 
TOF--PbSc and PbSc--PbSc pairs separately.
The systematic uncertainty associated with the embedding procedure was
estimated for the three centrality bins used in the yield determinations,
namely 0 - 10\%, 10 - 40\% and 40 - 92\%.
The systematic errors, calculated by estimating the occupancy efficiency
corrections for different track confirming criteria, were found to vary
from 7\% to 10\% for the three centrality bins used here, independent of the pair momenta.
Fig.~\ref{emb} shows the $\epsilon_{Occupancy}$
factors as a function of collision centrality
for the $K^{+}K^{-}$ pairs both identified in the TOF detector.
The occupancy dependent efficiency factors were found to be independent of
the transverse momenta of the pairs.
\section{Results and Discussion}
In this section, we present and discuss the results of our 
measurements, which consist of: 1) the $\phi$ line shape
analysis, 2) the transverse mass spectra analysis, 3) the integrated yields and ratios analysis, 
4) hydrodynamical fits to $\pi^{\pm}, K^{\pm}, p$, $\overline{p}$  
and $\phi$ transverse momentum spectra, and 5) the centrality 
dependence of the yields and  nuclear modification factor $R_{CP}$
as compared to that of pions and (anti)protons.
    
\subsection{Line Shape Analysis}
\label{sec:lineshape}
The invariant mass spectra of the $\phi$ mesons are obtained by subtracting
the combinatorial backgrounds from the same event $K^{+}K^{-}$ mass
spectra. The details
of the combinatorial background analysis were described in 
section III~D. For the best statistical precision, we combine data from 
TOF and PbSc detectors to analyze the $\phi$ mass centroids and widths
at the five different centrality bins.

\begin{figure}[t]
\includegraphics[width=1.0\linewidth]{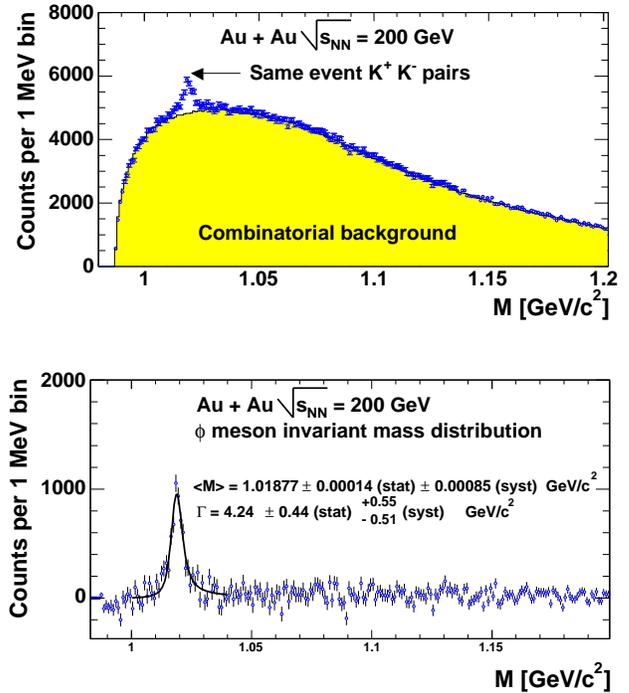}
\caption {\label{lineshape1} Minimum-bias $\phi \rightarrow K^{+}K^{-}$
invariant mass spectrum using the kaons identified in the
PHENIX detector. The top panel shows the same event (circles) and combinatorial background $K^{+}K^{-}$ mass distributions. The bottom panel shows the
subtracted mass spectrum fitted with Relativistic Breit-Wigner function convolved
with the Gaussian resolution function.}
\end{figure}

Fig.~\ref{lineshape1} shows the minimum-bias $\phi \rightarrow K^{+}K^{-}$ 
invariant mass spectrum for the PHENIX data. The subtracted
$\phi$ mass spectrum (lower panel), containing approximately 5100 $\phi$ in the
fit region,
is fitted with a relativistic Breit-Wigner (RBW) mass 
distribution function~\cite{Back03} convolved with a Gaussian experimental mass resolution function.
Using Monte Carlo studies based on the experimentally measured single kaon
momentum resolution, the experimental $\phi$ mass resolution
is calculated to be 1.0~MeV/$c^{2}$. 
This pair mass resolution value is found to be almost constant across the
kinematic region of acceptance.

The errors on the data points in Fig.~\ref{lineshape1} reflect the statistical errors only. 
The systematic errors associated with the mass centroid and 
width measurements originate from the
magnetic field uncertainties in the kaon momentum determination and the 
combinatorial background normalization procedure. The minimum-bias 
line shape parameters (centroid and width) derived in our analysis are listed in Table~\ref{lineshape2}.
The fitted minimum bias $\phi$~mass centroid and width are consistent at the
one standard deviation (1$\sigma$) level with the PDG values\footnote{
The PDG value for the $\phi$ mass centroid is $1019.456 \pm 0.020$~MeV/$c^2$,
and for the $\phi$ width the value is $4.26\pm 0.05$~MeV/$c^2$~\cite{PDG04}.},
taking into account both systematic and statistical errors.

\begin{table}
\caption{\label{lineshape2} $\phi$ meson mass centroid and width for the minimum-bias
Au + Au collisions at $\sqrt{s_{NN}}$ = 200 GeV.  The corresponding PDG values
are  $M_\phi = 1.019456$~GeV/$c^2$, $\Gamma_\phi = 4.26$~MeV/$c^2$}
\begin{ruledtabular}
\begin{tabular}{cc}
Parameter & Value \\\hline
Centroid &1.01877 $\pm$ 0.00014 (stat) $\pm$ 0.00085 (syst.) \\
(GeV/$c^{2}$) & \\
  & \\
Width & 4.24 $\pm$ 0.45 (stat) $\pm$ $^{0.55}_{0.51}$ (syst) \\
(MeV/$c^{2}$) & \\
\end{tabular}
\end{ruledtabular}
\end{table}

\begin{figure*}[htb]
\includegraphics[width=1.0\linewidth]{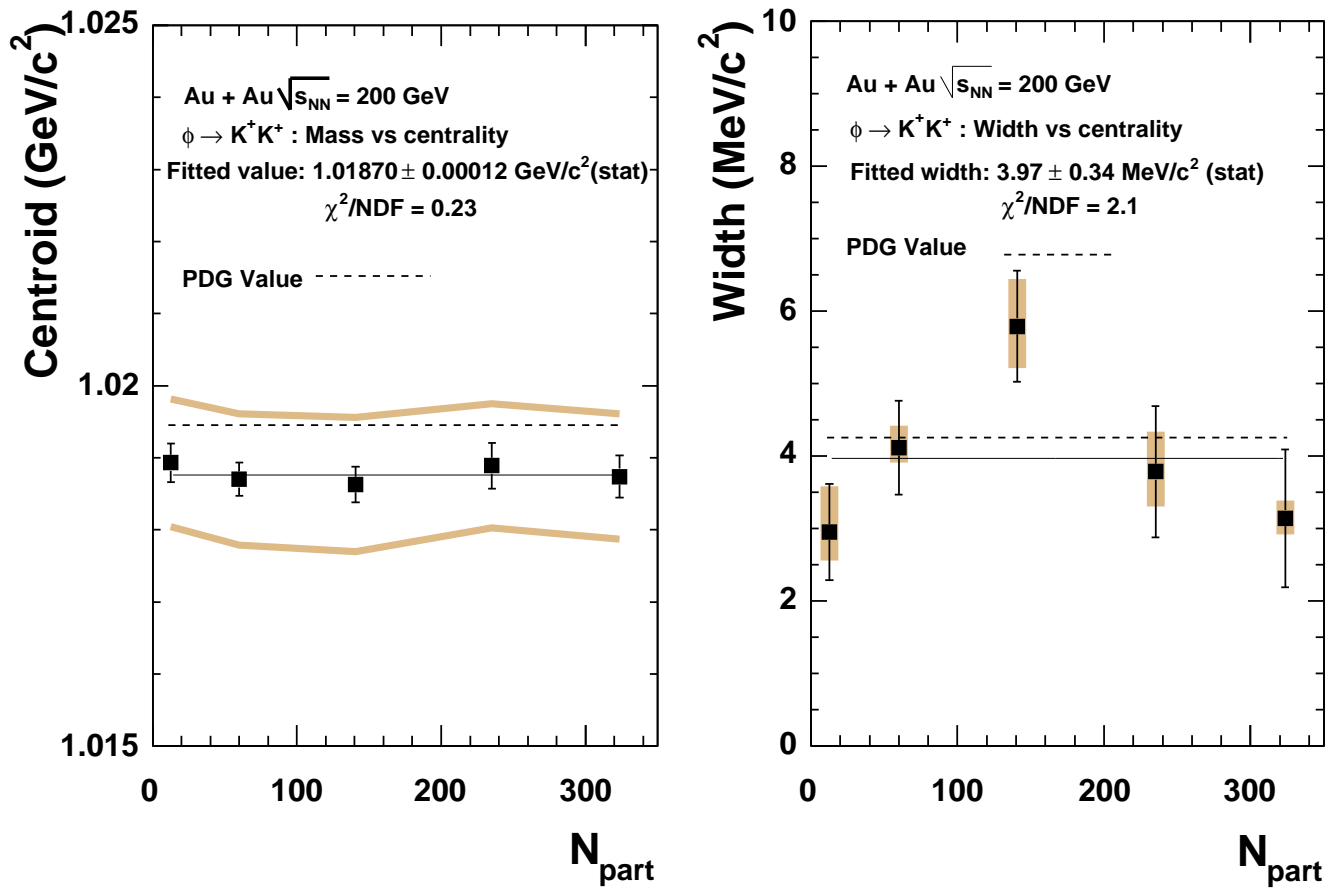}
\caption {\label{lineshape3} Centrality dependence of the $\phi$ mass
centroid (left) and $\phi$ intrinsic width (right), where the $N_{\hbox{part}}$
to Centrality correspondence is given in Table~\ref{ncoll}. For the
mass centroid plot, the 1$\sigma$~systematic error limits on the data points
are shown by the two continuous bands. The dotted line shows the PDG
centroid value (1.019456~GeV/$c^2$). The solid line indicates the centroid value obtained
from a one parameter fit assumption.  For the width plot
the systematic errors on the RBW widths are indicated as bands on each data
point. Similarly the dotted line shows the PDG width value (4.26~MeV/$c^2$),
and the solid line shows a
one parameter fit result for the measured data points.}
\end{figure*}

We investigated the centrality dependence of the $\phi$ meson line
shapes. For each 
centrality bin, we again fitted the $\phi$ mass spectrum with the
RBW function convolved with a Gaussian experimental $\phi$ mass resolution. The
results are presented in Fig.~\ref{lineshape3}. The left panel of the figure
shows the centrality dependence of the fitted centroids. The upper and lower
1$\sigma$ systematic error limits are indicated. The dotted line shows the
PDG mass centroid. The solid line indicates the result obtained
with a one--parameter constant fit through the
measured data points. 
These results lead to two immediate conclusions. First, to within less than 1~MeV/$c^2$ there is
no observed centrality dependence of the $\phi$ meson mass centroid, and second, the 
fitted centroids at all centralities are consistent with the PDG 
value within the statistical and systematic uncertainties of our measurements.

The $\phi$ mass widths, as shown in the right panel of the figure, are
studied as a function of the centrality. The error bar on
each point shows the statistical error while the bands on the points indicate
the systematic errors.  The dotted line shows the PDG $\phi$ mass 
width. The solid line shows the results of the constant fit assumption to the
data points. Again, within the random and systematic error limits shown, there
is no convincing evidence of a variation of the $\phi$~width as a function of centrality.

\begin{table*}
\caption{\label{phiPredict} Theoretical Predictions for Changes in the $\phi$ Resonance}
\begin{ruledtabular}
\begin{tabular}{cccc}
Authors & Models & Environment & Predictions \\ \hline
Caberara and Vacas~\cite{CabreraAndVacas03} &Chiral SU(3) & Cold hadronic &$M_\phi \rightarrow 1.011$~Gev\\
                                           &             &               &$\Gamma_\phi \rightarrow 30$~MeV/$c^2$\\
Pal, Ko, and Lin~\cite{Pal2002} & AMPT & Hot hadronic & $M_\phi \rightarrow
0.95$~GeV/$c^2$ at twice normal nuclear density ($\rho_0$)\\
                               & Chiral Lagrangian &      & $\Gamma_\phi \rightarrow 45$~MeV/$c^2$ at 2$\rho_0$\\
                               &                   &      & Suppression of $\phi\rightarrow K^+K^-$ relative to $\phi\rightarrow e^+e^-$\\ 
Oset and Ramos~\cite{OsetAndRamos01} & Kaon mass renormalization & Cold hadronic & $M_\phi$~unchanged \\
                                    &                           &               & $\Gamma_\phi \rightarrow 22$~MeV/$c^2$\\
Smith and Haglin~\cite{SmithAndHaglin98} & One boson exchange & Hot hadronic & $M_\phi$~unchanged \\
                                    &                         &               & $\Gamma_\phi \rightarrow 14\hbox{--}24$~MeV/$c^2$\\
Blaizot and Galain~\cite{BlaizotAndGalain91} & Nambu-Jona-Lasino & Hot hadronic &$M_\phi \rightarrow 2M_K$~at~$T\approx T_{\hbox{critical}}$ \\
                                    &                         &              &$\Gamma_\phi$ {\em reduced} by a factor of 6 \\
                                    &                         &              &$\phi \rightarrow K^+K^-$ disappears\\ 
Bi and Rafelski~\cite{BiAndRafelski91} & Bag Model         & Hot hadronic&$M_\phi \rightarrow 1.029$ GeV/$c^2$~at~$T\approx T_{\hbox{critical}}$ \\
                                      & Chiral Invariance &              &$\Gamma_\phi \rightarrow 10$~MeV/$c^2$ \\
\end{tabular}
\end{ruledtabular}
\end{table*}

The topic of medium effects on meson masses has been actively investigated
in the recent literature~\cite{Singh86, Singh87, Singh93, Singh97, SmithAndHaglin98, 
Klingl98, RappAndWambach00, AlvarezAndKoch02, CabreraAndVacas03}. The
predictions are that for both cold and hot nuclear matter there could
be a decrease of the $\phi$ mass value by a few~MeV/$c^2$ or even tens~of~MeV/$c^2$. Even more
dramatically the width could increase by several multiples above
the PDG value~of~4.26~MeV/$c^2$. A sample of such predictions is
given in Table~\ref{phiPredict}.  However, one
of the models~\cite{SmithAndHaglin98} considers the $\phi\rightarrow K^+K^-$
channel largely insensitive to medium effects since the kaons are
``unlikely to escape without reacting further, thus destroying
any useful information possessed about the $\phi$''. In this sense
the $\phi\rightarrow K^+K^-$ is inherently biased in that only
$\phi$ decays which are unaffected by the medium changes, for example
those produced peripherally, can be detected.

It is sometimes thought that since the vacuum $c\tau$ of the $\phi$
is $\approx 45$~fm/$c$ there will be only limited sensitivity to
medium effects in any case.  However, if the resonance width were
to actually increase by several times, as indicated in Table~\ref{phiPredict},
then the $c\tau$ would then
approach or be even substantially less than 10~fm/$c$, which is a size
compatible with the expected collision volume.  The dramatic
width changes predicted in either cold or hot nuclear matter
might be visible, at least in the dilepton channel if not also the
$K^+K^-$ channel, as a function of centrality.

The present mass centroid and width data, which are integrated over
the available \mt range, rule out any major changes with respect to
the PDG values. Specifically, the one parameter fit result of
$3.97 \pm 0.34$~MeV/$c^2$ obtained here excludes at the 99\% confidence
level a width value of 4.75~MeV/$c^2$ or greater.
Possibly at the lowest \mt values where
the $\phi$ would remain longer in the collision volume, or with
the availability of more finely binned centralities, there could
be visible evidence of in medium effects.  However, the current
data sample is insufficient to explore these possibilities.

It seems clear from the current set of the theoretical models that
an observed change in the $\phi$ width would not be itself indicative
of a QGP formation.  One would first have to constrain the cold nuclear
medium effects on the $\phi$ as could be obtained in d+Au collisions,
or by comparing peripheral Au+Au collision data results with
central collision data.  It is also important to measure
the $\phi$ mass in the dilepton $e^+e^-$ channel. That channel should be
more sensitive to the $\phi$ which are produced deeper or earlier
in the collision volume.
\subsection{Spectral Shapes Analysis}
At low $m_{T}$, the spectral shapes carry information about the kinetic freeze--out conditions.
Since the centrality 
dependence reveals the effect of the system size on the fireball evolution
it becomes desirable to study the 
centrality dependence of the spectral shapes. Transverse mass spectra were obtained in three centrality bins 
corresponding to  0--10\%, 10--40\% and
40--92\% of the total geometrical cross-section.
We count the same event $K^{+}K^{-}$ pairs within a defined
mass window ($\pm$ 5 MeV/$c^2$ with respect to the $\phi$ mass centroid) and
estimate the number of combinatorial background pairs within that window. In each centrality bin, the
data are divided into different $m_{T}$ bins. The invariant mass spectrum for the
same events and the background distributions are obtained for each of these $m_{T}$
bins. Finally,
the background is subtracted from the same event invariant mass spectrum
within the aforementioned $\pm 5$~MeV/$c^2$ $\phi$ mass window to determine
the number of reconstructed $\phi$ mesons within that $m_{T}$ bin. The reconstruction
of the $\phi$ in the Monte Carlo simulation takes into account the
effect of the $\phi$ mass window size.

\begin{figure}[ht]
\includegraphics[width=1.0\linewidth]{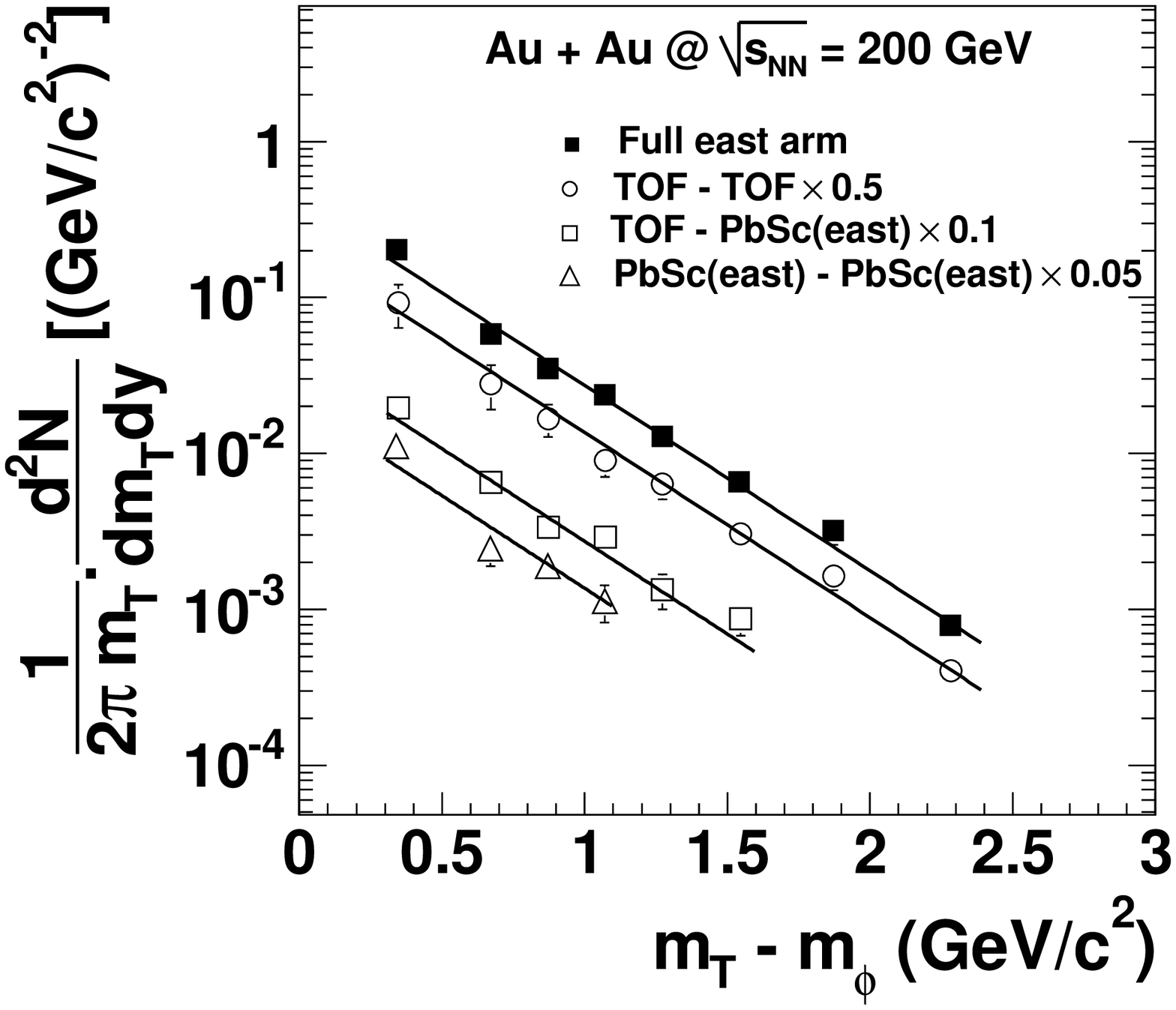}
\caption{\label{mt1} Minimum-bias $m_{T}$ spectra of the measured $\phi$ mesons
for three different PHENIX subsystem combinations, with scale factors as indicated. 
The combined spectrum is fitted with an exponential
function in $m_{T}$, Eq.~(3). The lines drawn through the individual spectra (TOF-TOF, TOF-PbSc and PbSc-PbSc) 
represent the same fit parameters as in the minimum-bias case. Statistical
error bars are shown.} 
\end{figure}

The $\phi$ mesons are reconstructed using kaons identified in the TOF and the 
PbSc detectors. Three detector combinations: TOF--TOF, TOF--PbSc,
and PbSc--PbSc are used to obtain three independent transverse mass spectra.
Fig.~\ref{mt1} shows the minimum-bias $m_{T}$ spectra for the above three combinations. The combined  
result, which is the sum of the three combinations is also included. For better visibility of the data points, 
TOF--TOF, TOF--PbSc and PbSc--PbSc  spectra are scaled by a factor of 0.5, 0.1 and 0.05, 
respectively. The $m_{T}$ spectrum obtained from the PHENIX detector is
fitted with the exponential function:
\begin{equation}
\frac{1}{2 \pi m_{T}} \frac{d^{2}N}{dm_{T}dy} = \frac{dN/dy}{2 \pi T (T + M_{\phi})} e^{-(m_{T} - m_{\phi})/T}
\end{equation}
where $dN/dy$ and the inverse slope $T$ are returned as two fitting
parameters. The lines drawn through the TOF--TOF, TOF--PbSc
and PbSc--PbSc spectra represent the same fit, but scaled with the same
scaling factors as the data points. 
Comparison of the individual spectra to the fit obtained from  the combined spectrum demonstrates the 
consistency between the different measurements, which have different systematic uncertainties.   
The TOF--TOF, TOF--PbSc and PbSc--PbSc spectra are also independently fitted
using Eq.~(3). The resulting 
$dN/dy$ and $T$ are tabulated in Table~\ref{mt01}. Both statistical and systematic errors 
are quoted. The systematic errors on $dN/dy$ originate from the systematic uncertainties associated with
extraction of the  yields in each $m_{T}$ bin (see Appendix~C) and the uncertainties from 
the fitting procedure. The latter is sensitive to the extrapolation of 
the $m_{T}$ spectra to $m_{T}=m_{\phi}$. A detailed account of 
systematic errors on $dN/dy$ and $T$ in the full dataset 
from all sources is shown in Appendix C.
We also fitted the TOF--TOF and TOF--PbSc data over the smaller $m_T$ range
of the PbSc--PbSc data and obtained consistent sets of $dN/dy$ and $T$ values from
that check.

\begin{table*}
\caption{\label{mt01} Minimum-bias $dN/dy$ and $T$ for different
subsystem combinations. The statistical and systematic errors are
shown after the first and second $\pm$ signs, respectively.}
\begin{ruledtabular}
\begin{tabular}{ccccc}
Subsystem combination&TOF-TOF&TOF-Pbsc&PbSc-PbSc& Full Data Set\\\hline
dN/dy&$1.16 \pm 0.17 \pm  0.19 $&$1.37 \pm 0.15 \pm 0.22 $&$1.47 \pm 0.26 \pm 0.27$&$1.34 \pm 0.09 \pm 0.21 $\\
T(MeV)&$380 \pm 18 \pm 22$&$385 \pm 34\pm 28$&$311 \pm 47\pm 65$&$366 \pm 11\pm 18$\\
\end{tabular}
\end{ruledtabular}
\end{table*}

Fig.~\ref{mt1} and Table~\ref{mt01} indicate that the three different analyses with different systematic 
uncertainties  give consistent results. This allowed us to combine the results and make use of the  maximum 
available statistics in each $m_{T}$ bin.  This combined spectrum 
was used to obtain the physics results discussed in the next sections.

\begin{figure}[ht]
\includegraphics[width=1.0\linewidth]{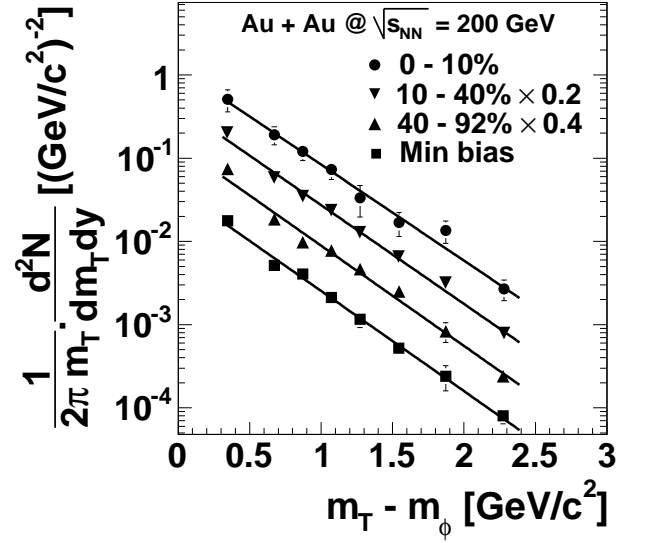}
 \caption{\label{mt2} $m_{T}$ spectra of $\phi$ mesons for 0--10\%, 10--40\%, 40--92\% and minimum-bias 
(0--92\%) centrality classes, with scale factors as indicated. Each spectrum is fitted with an exponential
function in $m_{T}$, Eq.~(3), with the fit parameters listed in Table~V.
Statistical error bars are shown.}
\end{figure}

\begin{figure}[ht]
\includegraphics[width=1.0\linewidth]{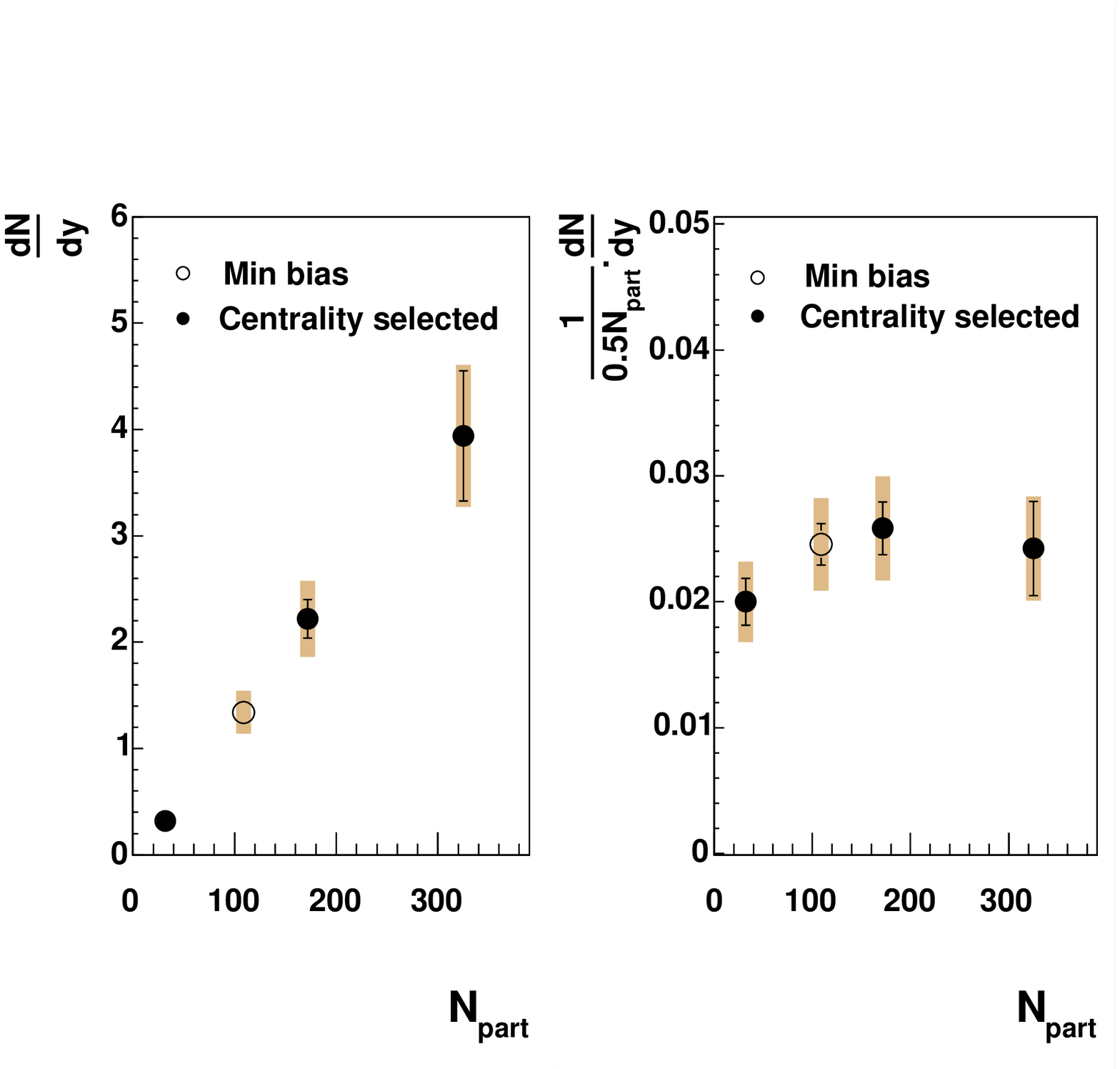}
   \caption{\label{cent1} Centrality dependence of $\phi$ yield at mid-rapidity. 
The value of $dN/dy$ increases steadily with
the number of participants ($N_{part}$)(left)
whereas the $dN/dy$ per participant pairs increases slightly
from peripheral to mid-central events and saturates after that (right).
The error bars indicate the statistical errors.
The shaded boxes on each data point are the systematic errors.
}
   \protect
\end{figure}

\begin{figure}[ht]
\includegraphics[width=1.0\linewidth]{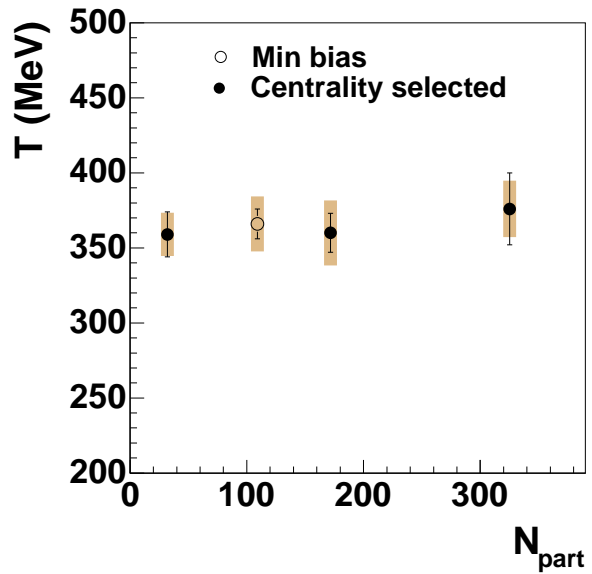}
   \caption{\label{cent2} Centrality dependence of the inverse slope, T.
}
   \protect
\end{figure}

Fig.~\ref{mt2} shows $m_{T}$ spectra of the $\phi$ mesons 
in 0--10\%, 10--40\%,  40--92\% and minimum bias
 centrality classes. The data points representing the invariant yield as a function
of transverse momentum are given in Appendix B. Each spectrum is fitted with an $m_{T}$-exponential 
function~Eq.~(3). The $\phi$ yield per unit of rapidity ($dN/dy$) and
inverse-slope ($T$) obtained from the fits are shown in Table~\ref{mt031},
and summarized in Fig.~\ref{cent1} and Fig.~\ref{cent2}
\footnote{The STAR experiment at RHIC has recently reported its analysis of the
$\phi\rightarrow K^+K^-$ data for Au+Au at
$\sqrt{s_{NN}}=200$~GeV\cite{adams04}. The analyses for PHENIX and STAR
have one common centrality bin, 0--10\%, for which the extracted $dN/dy$
are not in agreement.  The STAR value is $6.65\pm 0.35 (stat) \pm 0.73 (sys)$,
compared with the PHENIX value $3.94 \pm 0.60 (stat) \pm 0.62 (sys)$.  The
discrepancy persists even if one eliminates the four lowest $m_T$ data points
from the STAR data set in order to fit over the same $m_T$ range for both the
STAR and the PHENIX data.  There is not a discrepancy between the quoted
inverse slope parameters.}.

\begin{table*}
\caption{\label{mt031} $dN/dy$ and $T$ for different centrality bins.}
\begin{ruledtabular}
\begin{tabular}{ccc}
Centrality&dN/dy&T\\
(\%)&&(MeV)\\\hline
0 -- 10\%&3.94 $\pm$ 0.60 (stat) $\pm$ 0.62 (syst)&376 $\pm$ 24 (stat) $\pm$ 20 (syst)\\
10 -- 40\%&2.22 $\pm$ 0.18 (stat) $\pm$ 0.35 (syst)&360 $\pm$ 13 (stat) $\pm$ 23 (syst)\\
40 -- 92\%&0.32 $\pm$ 0.03 (stat) $\pm$ 0.05 (syst)&359 $\pm$ 15 (stat) $\pm$ 16 (syst)\\
Minimum Bias & 1.34 $\pm$ 0.09 (stat) $\pm$ 0.21 (syst) & 366 $\pm$ 11 (stat) $\pm$ 18 (syst)\\
\end{tabular}
\end{ruledtabular}
\end{table*}

\subsection{Yields and ratios}
\label{sec:ratios}

Hadron yields and ratios carry information about the chemical properties of the system.
The yields of strange particles have been of particular interest as they help in understanding the 
strangeness enhancement in heavy ion collisions and the equilibration of strangeness. 
It is important to study these phenomena as a function of system size. Centrality selected 
data can be particularly useful in this respect. Here we present the yield of the $\phi$ mesons 
at mid-rapidity as a function of centrality and compare this yield to the yields of other hadrons and 
the results from lower energy heavy ion collisions. 
The $dN/dy$ of $\phi$ (shown in Fig.~\ref{cent1}) is found to increase steadily with centrality.
In the right panel, the yield is normalized to the number of participant pairs to take into account 
the size of the system. Within the error bars this normalized rapidity density is approximately 
independent of centrality with a possible slight increase from peripheral to the mid-peripheral collisions.
The trend is quite different from lower energy results measured at the  
AGS. In~\cite{Back03}, the yield of $\phi$ was reported to be
increasing faster than linearly with the number of participants.

\begin{figure}[ht]
\includegraphics[width=1.0\linewidth]{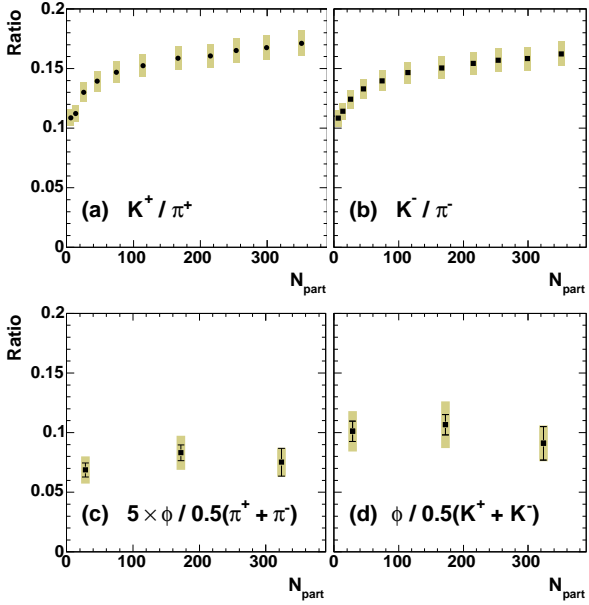}
\caption{\label{strange-npart} Centrality dependence of particle ratios
for (a) $K^{+} / \pi^{+}$, (b) $K^{-} / \pi^{-}$, (c) $\phi / 0.5
~(\pi^{+}+\pi^{-})$ (scaled by a factor of 5),
 (d)  $\phi / 0.5~(K^{+}+K^{-})$ in Au+Au collisions at $\sqrt{s_{NN}}$ = 200
GeV.}
\end{figure}

We now consider the ratio of strange to non-strange particles
in order to understand the extent and mechanism of the strangeness 
enhancement in heavy ion collisions. The ratios $K^{+} / \pi^{+} $ and  $K^{-} / \pi^{-}$, 
are shown in Fig.~\ref{strange-npart} 
(a)--(b). Both ratios show an increase of $\approx 60\%$ from peripheral to central collisions.
Most of this increase is for $N_{part} < 100$. Only a mild increase or saturation is observed 
from mid-central to the top centrality bin~\cite{ppg026}. Fig.~\ref{strange-npart} (c)--(d)
shows the centrality dependence of the $\phi / \pi$ and $\phi / K$ ratios.
The limited statistics prevent us from extending the $\phi$ measurements into 
larger number of centrality bins spanning the more peripheral events.
The $\phi / K$ ratio, in this limited number of centrality bins,
is approximately flat as a function of centrality. 
The possibility of structure in the $\phi/\pi$ ratio is difficult
to infer from our data within the error bars.  One might expect to see
some centrality dependence in the $\phi/\pi$ ration because there is obviously a
centrality dependence to the $K/\pi$ ratio. However, we do not
have enough centrality bins, nor enough signal in each bin,
from which to conclusively identify a centrality dependence
in $\phi/\pi$. The flat behavior of the $\phi/K$ ratio
indicates that there is no pronounced difference in the production of open (kaon)
and hidden ($\phi$) strangeness in heavy ion collisions at RHIC energies. Production in the 
hadronic stage via kaon coalescence $K^+K^- \rightarrow \phi$, seems to be excluded by that data, 
as it would result in an increase in   $\phi/K^{-}$ as a function of $N_{part}$, which is not observed.   

\begin{figure}[ht]
\includegraphics[width=1.0\linewidth]{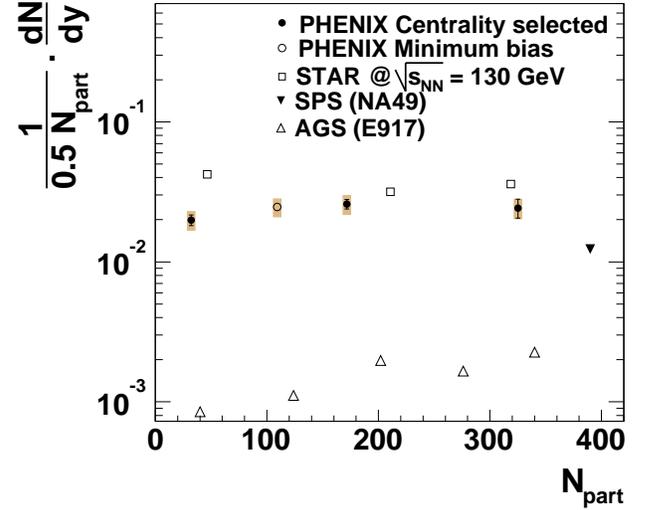}
\caption{\label{coll} Centrality dependence of $\phi$ yields at different
collision energies. STAR data are from \cite{Adler02},
NA49 data are from \cite{Afanasiev00}, and E917 are from \cite{Back03}.
}
\end{figure}

The collision energy dependence of the $\phi$ yield is shown in
Fig.~\ref{coll} where we plot the $dN/dy$ per participant pair as a function
of the number of participants for different colliding energies. 
The figure indicates two aspects of $\phi$ meson
production at various energies. First, as we go from AGS to SPS to RHIC,
the $\phi$ meson yield per participant increases by an order of magnitude
overall. Secondly, at the
AGS energy, we find a steady increase of $\phi$ production per
participant pair from peripheral to central collisions. It is worth mentioning
that the NA50 experiment~\cite{NA50} at CERN SPS  reported an increase
in fiducial $\phi$ yield (in  $\mu^{+} ~\mu^{-}$ decay channel) per
 participant from peripheral to central collisions although the yield per participant showed
saturation within error for the  top centrality bins.
The yield of $\phi$ mesons at RHIC, on the contrary, 
is found to be almost independent of centrality.

\begin{figure}[ht]
\includegraphics[width=1.0\linewidth]{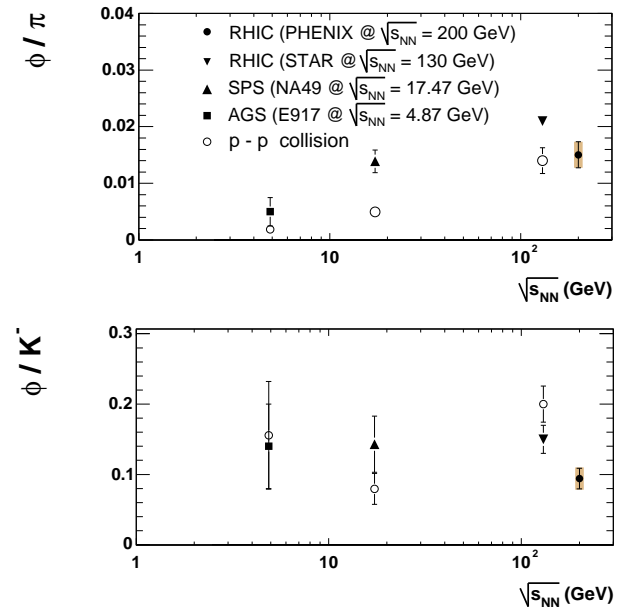}
\caption{\label{coll2} $\phi/\pi$ (top half) and $\phi/K^{-}$ (bottom half)
ratios as a function of collision energy.  The data symbols have the same
meanings in both the top and the bottom halves of the figure.
}
\end{figure}

To investigate further the mechanism of $\phi$ enhancement with
increase in collision energy, we study the two ratios $\phi/\pi$ and $\phi/K^{-}$
as a function of collision energy as illustrated in Fig.~\ref{coll2}. The 
$\phi/\pi$ ratio is found to increase with
the collision energy from AGS to RHIC.
The $\phi/K^{-}$ ratio, on the other hand, remains almost constant 
within error bar with increasing collision energy.

\subsection{Hydrodynamical Model Fits to the  Spectral Shape Data}
\label{sec:blastwave}


From the $\phi$ spectral data shown in the preceding sections, we can conclude
that the transverse mass distributions are well described by an exponential 
distribution and are quite similar for all the centralities.  There is little, if any, 
centrality dependence of the inverse slope parameter in the measured centrality bins, as
shown in Fig.~\ref{cent2}. The exponential behavior is expected for particle production 
from a thermal source. 

If the system develops collective motion, particles experience
a velocity boost resulting in an additional
transverse kinetic energy component. This motivates the use of 
the transverse kinetic energy, {\it i.e.}~transverse mass minus the 
particle rest mass, for studying flow effects. Traditionally, the CERN 
experiments~\cite{NA44_mtSlopes,WA97_mtSlopes} have used simple exponential 
fits to the transverse kinetic energy distributions and often quote just one number, 
the inverse slope $T$, to characterize the spectra. These fits are usually done in the  
the range $(m_T-m_0) < 1$ GeV/$c^2$ in order to minimize the contribution from hard processes.

\begin{figure}[ht]
\includegraphics[width=1.0\linewidth]{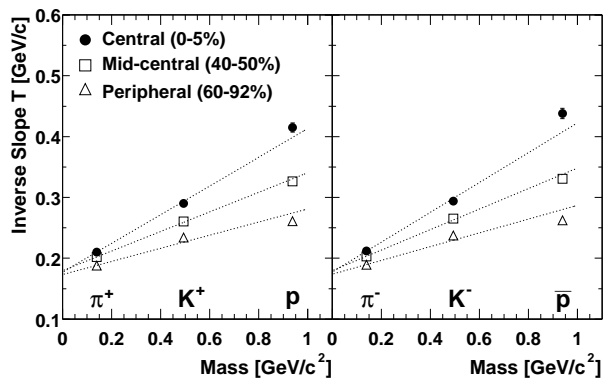}
\caption{Mass and centrality dependence of inverse slope parameters $T$
in $m_T$ spectra for positive (left) and negative (right) particles 
in Au+Au collisions at $\sqrt{s_{NN}}$~=~200~GeV. The fit ranges are
0.2 -- 1.0~GeV/$c^{2}$ for pions and 0.1 -- 1.0~GeV/$c^{2}$ for kaons, 
protons, and anti-protons in $m_T - m_0$. Figure taken from reference~\cite{ppg026}.}
\label{fig:slope_mass}
\end{figure}

The results of such fits, obtained from previously published PHENIX $\pi^{\pm},K^{\pm},p$ and $\overline{p}$
data~\cite{ppg026} are shown in Fig.~\ref{fig:slope_mass}. The slope parameters show a clear mass dependence, 
as expected from radial flow. The mass dependence increases from peripheral to central collisions indicating 
stronger collectivity in the more central events. The $\phi$ meson has mass similar to that of the proton. 
Hence we expect that if the $\phi$ participates in the collective flow
then its inverse slope will be affected by this 
motion. For protons, the slope parameter changes significantly from peripheral to central
collisions. As just noted, the $\phi$ inverse slopes shows no such centrality dependence. 
An important difference between the results obtained for $p,\overline{p}$ and $\phi$ is that the former have been fitted within a limited 
low-$m_{T}$ range ( $(m_T-m_0) < 1$ GeV/$c^2$) as motivated above. In the case
of the $\phi$ the full measured 
range has been used for the fit due to having limited data at low--$m_{T}$.
As shown in Fig.~\ref{mt2}, the three data points below 1~GeV/$c^2$ are consistent with
the fit over the entire m$_T$ range.

The detailed study of the higher statistics (anti)proton spectra~\cite{ppg026} indicated that
(anti)proton spectra cannot
be described by a single exponential in $m_{T} - m_{0}$, if the full measured
range is considered. 
Although easy to visualize, the one-parameter
inverse slope analysis proves to be too simplistic as a way to infer the
kinetic
properties of the system. In particular, the degree to which the $\phi$ mesons participate in the
collective expansion will be shown (see Fig.~\ref{ptspectra}) to be obscured
in this simple approach.


A more sophisticated approach to this problem is to compare the particle spectra to a functional form which 
describes a boosted thermal source, based on relativistic hydrodynamics~\cite{sollfrank}.
This is a two--parameter model, termed the ``blast--wave'' model, in which the surface radial flow 
velocity ($\beta_{T}$) and the freeze-out temperature ($T_{fo}$) are extracted from the 
invariant cross section data according to the equation 
\begin{equation}
{{dN}\over{m_T\,dm_T}} \propto
	\int_{0}^{R} f(r)\,r\,dr \, m_T 
	I_0\Big(\frac{p_T \sinh\rho}{T_{fo}}\Big)
	K_1\Big(\frac{m_T \cosh\rho}{T_{fo}}\Big),
\end{equation}
where $I_0$ and $K_1$ represent modified Bessel functions with
$\rho$ being the transverse boost which depends on the radial position according to 
\begin{equation}
\rho = \tanh^{-1}(\beta_{T}) \cdot r/R
\end{equation}
Here the parameter $R$ is the maximum radius of the expanding 
source at freeze-out.  The function $f(r)$ represents the density
which is taken to be uniform in this calculation.

To study the parameter correlations, we make a grid of
($T_{fo}$, $\beta_T$) pairs and then for each pair we perform a chi-squared minimization 
for each particle type.  We use a linear velocity profile and 
constant particle density distribution. The first fit attempt is performed simultaneously for 
the six particle species $\pi^{\pm}, K^{\pm}$, and
the $p,\overline{p}$ in the range $(m_T - m_0) < $ 1.0 GeV/$c^2$.

The experimental data for $p$ and $ \overline{p}$ have been corrected for 
$\Lambda$ and $\overline{\Lambda}$ decays.
However, the invariant yields of $\pi^{\pm}$ and $K^{\pm}$  
include feed--down from  the decay of resonances and weak decays. 
To take this into account we add the decay of mesonic ($\rho, \eta, \omega, K^{*}$...) and 
baryonic ($\Delta,\Lambda,\Sigma$...) resonances as follows:
\begin{itemize}
\item [1)] Generate resonances with the transverse momentum distribution
determined by each combination of $T_{fo}$ and $\beta_T$.

\item [2)] Simulate the decays  using a Monte Carlo approach 
and obtain $\pi^{\pm}$ and  $K^{\pm}$ distributions.

\item [3)] Merge all particles, where the particle abundance
is calculated with chemical parameters~\cite{munzinger}
$T_{ch}$~=~177~MeV, $\mu_{B}$~=~29~MeV.
\end{itemize}

\begin{figure}[ht]
\includegraphics[width=1.0\linewidth]{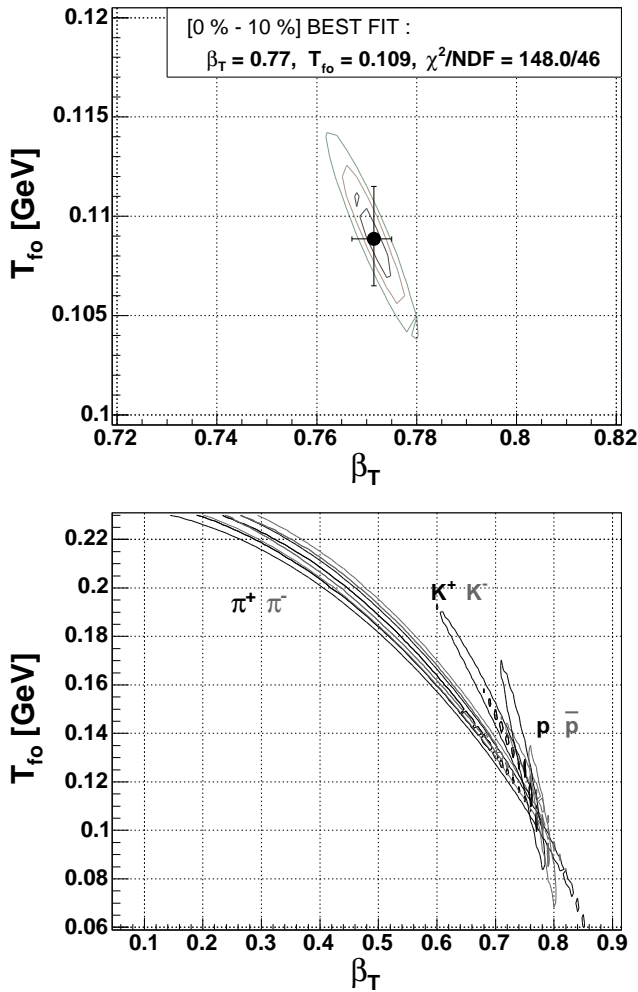}
   \caption{\label{fg:hydfit6} Contour plots for the hydrodynamical fit
to the 200~GeV single particle transverse momentum spectra for the
$\pi^{\pm}, K^{\pm}$, and $p,\overline p$ in the 0--10\% centrality bin.
The contour lines are in one standard deviation steps.
The upper plot is from a simultaneous
fit with the best value shown as the dot.  The lower plot is from independent
fits for the six particle spectra.}
   \protect
\end{figure}

\begin{figure*}[ht]
\includegraphics[width=1.0\linewidth]{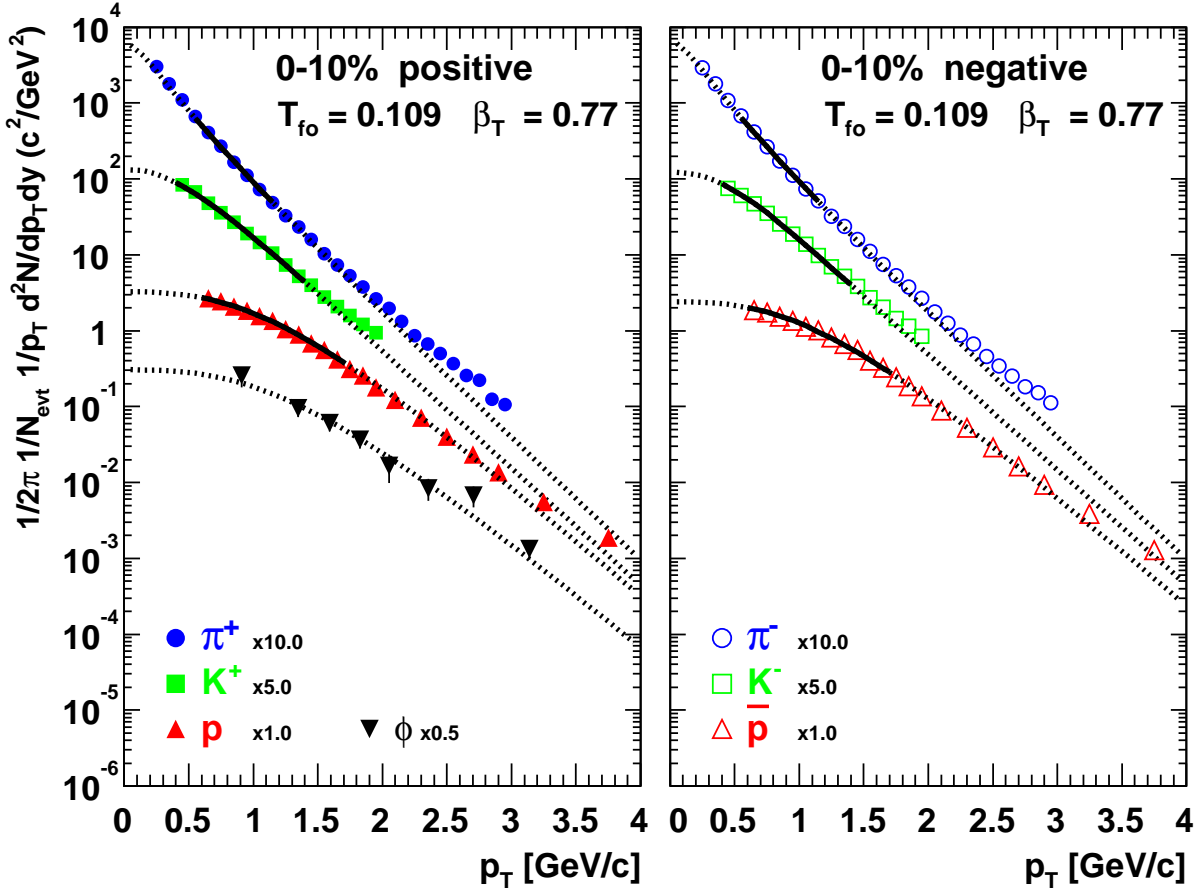}
   \caption{\label{fg:simfit6_0-10} Transverse momentum data and
best fit hydrodynamical results for the 0--10\% centrality bin
for the 200 GeV  $\pi^{\pm}, K^{\pm}$, and $p,\overline p$,
along with the prediction for the $\phi$ transverse momentum spectrum.
The transverse momentum ranges for the fit are indicated by the solid
lines, while the dashed lines indicated the extrapolated predictions
for each particle species data.
}
   \protect
\end{figure*}

The two--parameter $T_{fo}$ vs $\beta_T$ 
fit results obtained in this analysis for the most central bin
are shown in Fig.~\ref{fg:hydfit6}. Shown in the lower panel of the 
figure are the $\chi^{2}$ contour levels obtained from fitting  each particle spectrum separately. 
We observe that the parameters  $T_{fo}$ and $\beta_T$ are anti-correlated, the different 
particles have different preferred parameter space and different sensitivity to the parameters. 
For example, the contours for the heavier particles are more sensitive to the flow velocity than to the
kinetic freeze-out temperature. The minimum valleys in the
contours for the six particle species do overlap at a single common point at the 2$\sigma$ level.
To find the values of the parameters at this overlap point, a simultaneous fit for the
six single particle spectra ($\pi^{\pm}, K^{\pm}$, and $p,\overline p$) was done which
converges to a best fit value of 
$T_{fo}=108.9_{-2.4}^{+2.6}$~MeV and $\beta_{T} = 0.771_{-0.004}^{+0.003}$. Using
these parameters, we obtain the transverse momentum shapes shown
in Fig.~\ref{fg:simfit6_0-10} where we also include the prediction
for the $\phi$ spectrum shape which was not part of the original fit.
The shape of the $\phi$ spectrum is reproduced well.

\begin{figure*}[ht]
\includegraphics[width=1.0\linewidth]{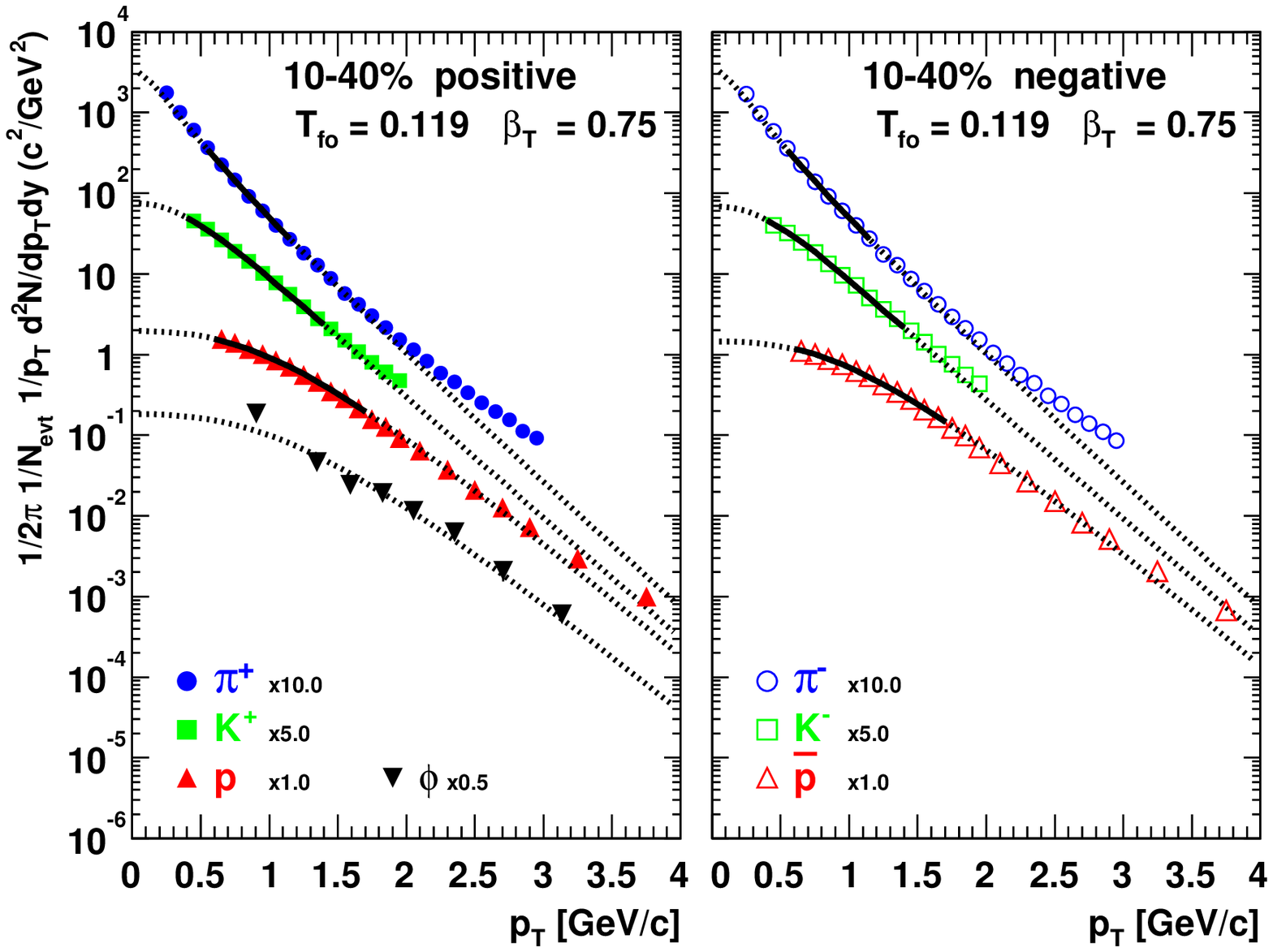}
   \caption{\label{fg:simfit6_10-40} Transverse momentum data and
best fit hydrodynamical results for the 10--40\% centrality bin
for the 200 GeV  $\pi^{\pm}, K^{\pm}$, and $p,\overline p$,
along with the prediction for the $\phi$ transverse momentum spectrum.
}
   \protect
\end{figure*}

\begin{figure*}[ht]
\includegraphics[width=1.0\linewidth]{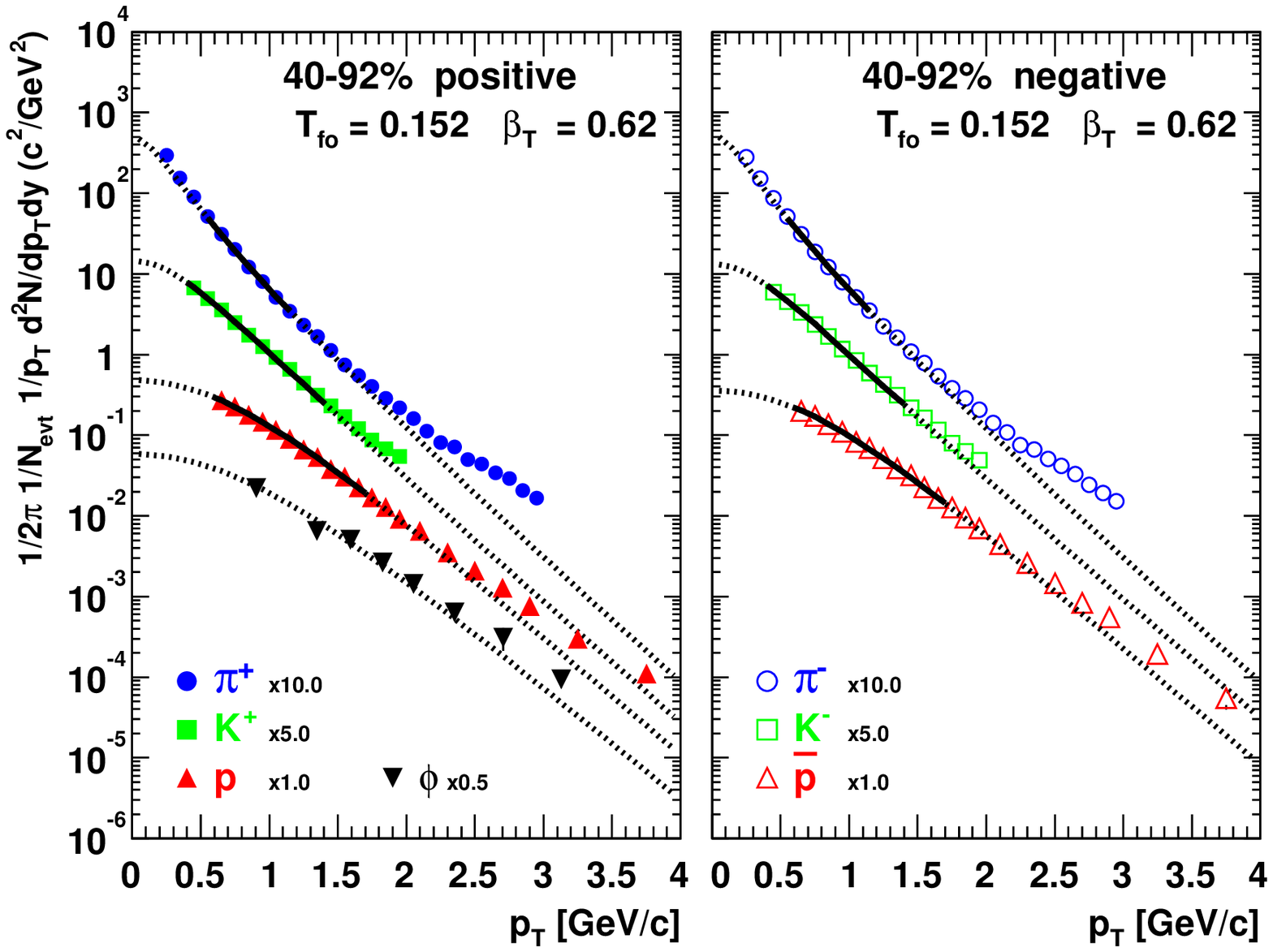}
   \caption{\label{fg:simfit6_40-92} Transverse momentum data and
best fit hydrodynamical results for the 40--92\% centrality bin
for the 200 GeV  $\pi^{\pm}, K^{\pm}$, and $p,\overline p$,
along with the prediction for the $\phi$ transverse momentum spectrum.
}
   \protect
\end{figure*}

For the two other centrality bins in this study, 10--40\% and
40--92\%, we show the best fit hydrodynamical results in
Fig.~\ref{fg:simfit6_10-40} and Fig.~\ref{fg:simfit6_40-92}, respectively.
Again we see that the $\phi$ transverse momentum shapes are reproduced
by the same flow parameters which fit the identified hadron data
at the same centrality bins.

For a second hydrodynamical fit attempt, we include the $\phi$
transverse momentum 0--10\% centrality data along with the previously identified hadron
data as part of the $\chi^2$~minimization search. 
The flow parameters derived with the $\phi$ data included
are numerically consistent with the flow parameters derived previously
without the $\phi$ data being included.

The two--dimensional grid search best fit values for the blast--wave
parameterization as a function of centrality are tabulated in
Table~\ref{blastWaveParam}. The radial average expansion velocity
$<\beta_T>$ is also given in this table.
For the range of centralities studied here,
the expansion velocity parameter is seen to decrease moderately for more
peripheral collisions while the kinetic freeze-out temperature increases more
significantly, approximately~$40\%$.  If one takes these parameters literally,
then the more peripheral collisions are subject to decreased radial flow
while correspondingly the particles are decoupling kinetically from
each other at temperatures closer to the chemical freeze-out temperature.
This is a physically reasonable scenario given fewer participants in
the initial expansion phase.

It should be pointed out that our present $\phi$ transverse momentum
range does not extend below 0.8~GeV/$c$. The spectral shapes at low-$m_{T}$, especially for the heavier 
particles, are mostly sensitive to the expansion velocity. In the range of our $\phi$ measurement, it is more 
appropriate to consider the asymptotic behavior of the spectral shapes, which for  $m_{T}>>m_{0}$ is given 
by~\cite{sollfrank}:
\begin{equation}
T_{eff} = T_{fo} \sqrt{ (1+\beta_{T})/(1-\beta_{T})}
\label{hydro:asympt}
\end{equation}
Here, $T_{eff}$ is the slope parameter obtained 
using $m_{T}$ exponential fit, as in Eq.~(3).  
It is interesting to note that the measured asymptotic 
slopes do not seem to depend on centrality, although both  $T_{fo}$ and $\beta_{T}$ show a clear centrality 
dependence. This is either due to a cancellation effect in Eq.~(\ref{hydro:asympt}), since the parameters 
are anti-correlated, or indicates that the hydrodynamics description is no longer valid at these large 
transverse momenta. We conclude that although  the $\phi$ data themselves cannot constrain the kinematic
 freeze--out conditions, they are consistent with the  hydrodynamical results obtained from the simultaneous 
fit to the $\pi^{\pm}, K^{\pm}, p $ and $\overline{p}$ spectra.


\begin{table}[h]
\begin{center}
\caption{\label{blastWaveParam} Blast wave model parameters~\cite{sollfrank} as a
function of centrality from fitting
$\pi^{\pm},K^{\pm}, p$ and $\overline{p}$ spectra.
The fit parameters quoted here are the results from fitting the six identified hadrons spectra
 simultaneously, without including the $\phi$.}
\begin{ruledtabular}
\begin{tabular}{rcccc}

Centrality & $T_{fo}$ [MeV]   &   $\beta_{T}$   & $<\beta_{T}>$ & $\chi^{2}/NDF$ \\ \hline

 0--10\% & $108.9_{-2.4}^{+2.6}$   & $0.771_{-0.004}^{+0.003}$   & $0.572_{-0.003}^{+0.003}$ & 148.0/46 \\
10--40\% & $119.0_{-1.5}^{+1.5}$   & $0.748_{-0.003}^{+0.003}$   & $0.550_{-0.002}^{+0.002}$  & 212.1/46 \\
40--92\% & $150_{-2}^{+2}$   & $0.630_{-0.005}^{+0.005}$    & $0.447_{-0.004}^{+0.004}$  & 150.9/46 \\
\end{tabular}
\end{ruledtabular}
\end{center}
\end{table}

\subsection{Nuclear modification factor $R_{CP}$ for $\phi$ mesons}
\label{sec:Rcp}
\begin{figure}[ht]
\includegraphics[width=1.0\linewidth]{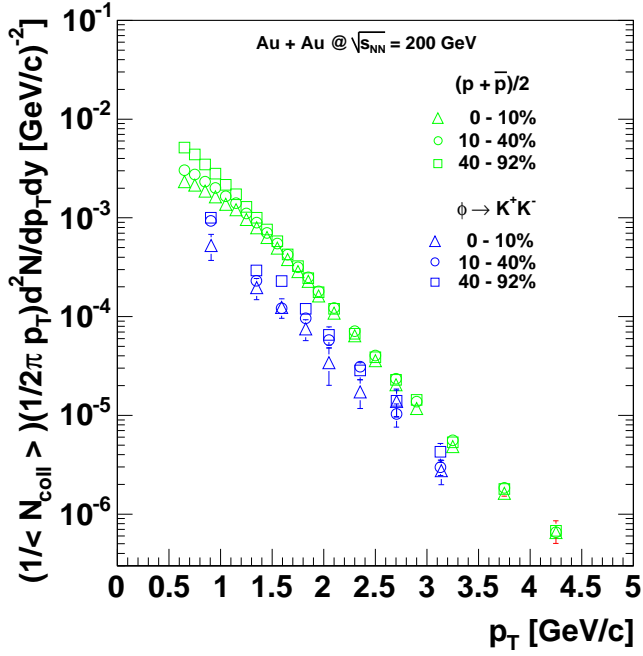}
   \caption{\label{ptspectra} $p_{T}$ spectra of proton and $\phi$ mesons
at different centralities scaled down by their respective number of $N_{coll}$
}
   \protect
\end{figure}
\begin{figure*}[ht]
\includegraphics[width=1.0\linewidth]{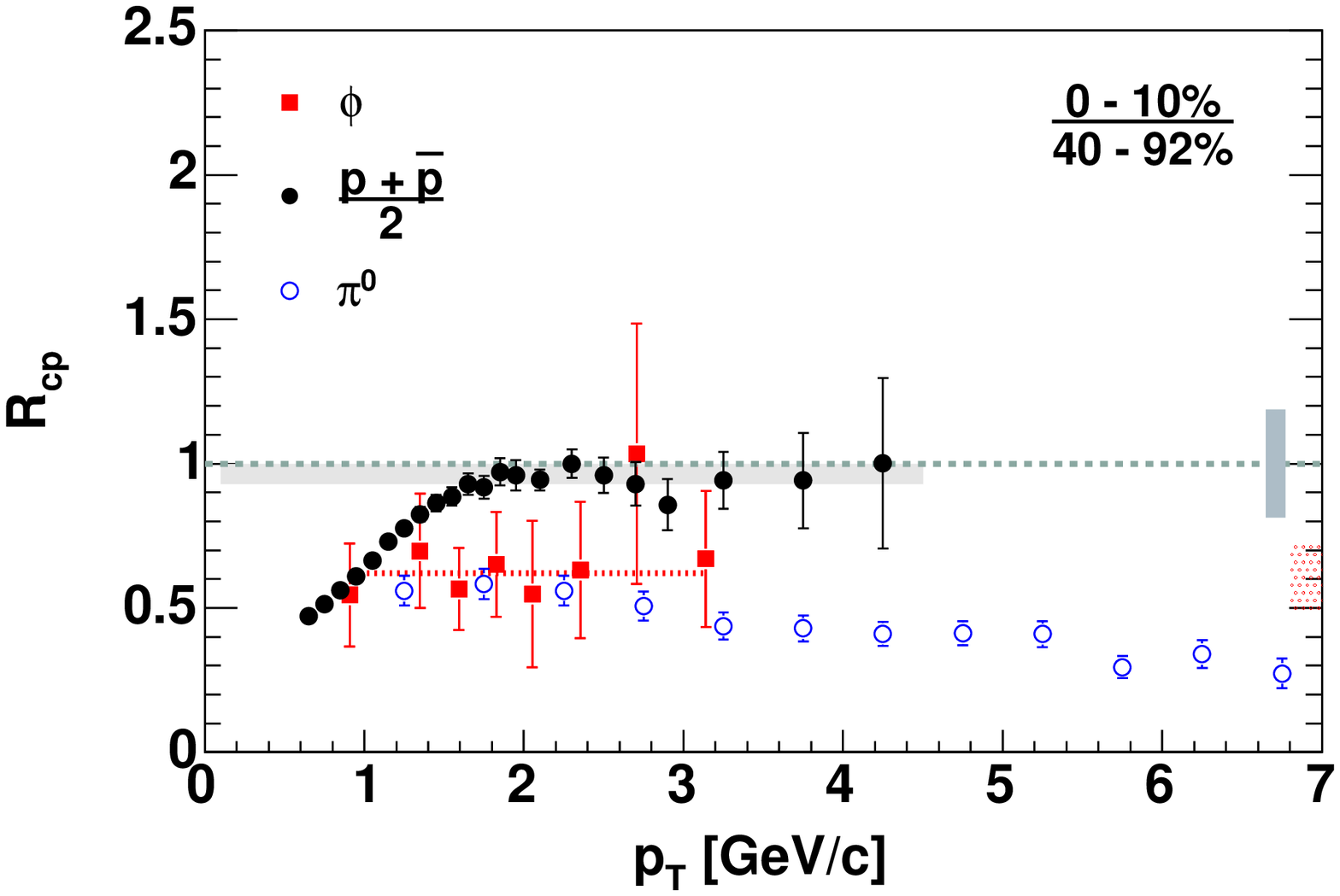}
\caption{\label{rcp} $N_{coll}$ scaled central to
   peripheral ratio $R_{CP}$ for $(p + \overline{p})$/2, $\pi^0$, and
   $\phi$. The proton and pion results are published~\cite{protonscaling}.
   The vertical dotted bar on the right represents the error on $N^{0-10\%}_{coll}
   /N^{40-92\%}_{coll}$ .  The shaded solid bar around RCP = 1 represents
   12\% systematic error which can move the proton and/or $\phi$ points
   with respect to one another.  The dotted horizontal line at $R_{CP}$ = 0.62 is a
   straight line fit to the $\phi$ data.}
\protect
\end{figure*}
One of the most important result demonstrated so far 
in the study of relativistic heavy ions at
RHIC is the observed suppression of high \pt pions in central
collisions as compared to either $pp$ collisions or peripheral Au~+~Au
collisions~\cite{Adler03}. A second, quite surprising observation is the unusually
large (anti)proton-to-pion ratio at high \pt . In particular PHENIX has
observed~\cite{protonscaling} that in central Au~+~Au collisions the
 $p/\pi$ and $\overline{p}/\pi$ ratios are
enhanced by a factor of 3 at intermediate $p_T$ ($1.5<p_T<4.5$ GeV/$c$)
as compared to the ratios in $pp$ collisions and the ratios 
obtained in quark and gluon jets measured in $e^+e^-$ collisions~\cite{DELPHI}.
It was also observed that proton and anti-proton production scales with
$N_{coll}$ in this $p_T$ region, in sharp contrast to the
strong suppression of pion production~\cite{protonscaling}. In $pp$  
collisions high \pt particles are the result of the
fragmentation of partons. Because of the power law nature of the hard
scattering spectrum, most of the particles at high \pt are expected
to be leading hadrons. The fragmentation functions, at least in the
vacuum, are expected to be universal and independent of the colliding
system under consideration. However, at intermediate $p_{T}$~(1.5--4.5 GeV/$c$)
 the PHENIX results from central Au~+~Au collisions 
are inconsistent with  the known fragmentation functions.

There are several conjectures which may explain the unexpected PHENIX result:
\begin{itemize}
\item [1)] hydrodynamic flow generated from the hadronic stage~\cite{flow}, or
\item [2)] hydrodynamic flow generated at a partonic stage together with
particle production from the recombination\cite{Fries03a, Hwa:2004ng, Greco:2003mm}
\item [3)] baryon junctions as a mechanism for an usually large build up of
baryons and anti-baryons at mid-rapidity~\cite{bjunctions},
\item [4)] in-medium modified fragmentation functions~\cite{xnwang}.
\end{itemize}

The first three possibilities invoke soft processes to populate a region of $p_{T}$ 
that is dominated by hard-scattering in $pp$ collisions. The soft production at intermediate $p_{T}$ 
is enhanced for protons and anti-protons, while pions remain dominated by hard-scattering.  In conjecture (4), 
the production for both pions and $p,\overline{p}$ is by hard-scattering, but the fragmentation 
functions are modified in the medium, thus resulting in an enhanced $p/\pi$ and $\overline{p}/\pi$ ratios. 
      
The second of these conjectures is 
particularly important since this hypothesis presupposes a
partonic state with collective behavior.  A critical factor
which may differentiate among these scenarios is whether the large
abundance of protons is due to its mass, or to the number of
constituent valence quarks~\cite{Fries03a, Hwa:2004ng, Greco:2003mm,
Greco:2003xt}. 
Hydrodynamic flow generated at the
hadronic stage imparts a single velocity to the moving matter,
hence similar mass particles should exhibit the same
momentum increase from this effect. In contrast, hypotheses
2 and 3 are dependent on whether the particles are baryons
or mesons.

The $\phi$ meson which has a mass similar to that of
a proton yet, like the pion, has two valence quarks, should distinguish
between (1) and (2) or (3). We examine the scaling properties of the intermediate $p_{T}$ yields  
of the $\phi$ and compare those to the yields of $(p+\overline{p})/2$. Fig.~\ref{ptspectra} shows
the transverse momentum spectra measured in three different centralities, each scaled down by the 
number of binary nucleon-nucleon collisions. The (anti)proton spectra show two pronounced features.
Below $p_{T} <1.5$~GeV/$c$, the spectral shapes are strongly influenced by the radial flow and 
thus the more central data have harder slope. Above $p_{T} = 1.5$ GeV/$c$, the effect of radial flow is 
negligible. The spectra converge to the same line.  Moreover, they scale with $N_{coll}$ for all centrality 
classes, as expected for hard-scattering unaffected by the nuclear medium. The $\phi$ spectra have a 
quite different behavior. There is no visible curvature at lower $p_{T}$ since 
this curvature is not expected to be prominent in the measured region.
At higher $p_{T}$ the $\phi$ spectra run parallel to the (anti)proton spectra, but do not 
obey $N_{coll}$ scaling.  To examine this feature on a linear scale,
we plot the ratio between the central and peripheral data,
{\it i.e.}~the ratio $R_{CP}$ (Fig.~\ref{rcp}).  
The systematic error arising from the determination of $N_{coll}$ is
represented by the dotted bar and is about 19\%.  This systematic error is 
common for all three particle species shown in the figure. The solid bar around $R_{CP} = 1$ 
represents the  $N_{coll}$ error for the protons (the same 19\%). Since these errors are correlated,
if the curves in the figure were to change due to a change in the value of $N_{coll}$, they will move 
together within the extent of the error bars shown. 
We would like to emphasize the  {\it comparison\rm} between the $R_{CP}$ values for the protons 
and the $\phi$. In this comparison, the systematic errors in determining $N_{coll}$ cancel.
The important systematic errors to consider are those that can move 
the $\phi$ points with respect to the proton points. When determining a 
ratio of spectra measured at different centralities, most systematic errors cancel. 
After removing the $N_{coll}$ error, the sources of error that remain for the $\phi$ come from 
the multiplicity dependent corrections and the effect of the  mass window as described above. 
For the protons and pions~\cite{ppg014,protonscaling}, the error that remains that is independent from 
the error determined for the $\phi$, comes from the multiplicity dependent corrections. 
The relative error between the $\phi$ and the proton measurements is evaluated at 7\% and is 
represented by the extended solid bar just below $R_{CP}=1$. Clearly the $\phi$'s behavior is more 
like that of the pions and not like that of the protons.  Thus we conclude that the $\phi$ meson exhibits
a suppression effect at intermediate $p_T$ similar to that of the pions. Although we can
not conclude whether $\phi$ production  at this intermediate $p_{T}$ is dominated by soft or hard 
processes, this observation provides support for models which depend on the number of valence quarks in the
particle as opposed to models which depend upon just the mass of the particle in order to 
explain the anomalous proton yields.
\section{Conclusions}
A systematic measurement has been made
of $\phi$ production at mid-rapidity in Au+Au collisions with $\sqrt{s_{NN}}=200$~GeV
at RHIC. The general features of the data include the yield
which rises 
from 0.318$\pm$0.028(stat)$\pm$0.051(syst) in peripheral collisions 
to 3.94$\pm$0.60(stat)$\pm$0.62(syst) in central collisions. There is seen to be little
centrality dependence to the inverse slope which is about 360~MeV.  The centroid mass and
resonance width are extracted with high enough precision to rule out any large (MeV/$c^2$ scale)
deviations from the accepted PDG values.
At \pt below 1.5 GeV/$c$, a blast wave analysis of the most central pion,
kaon and proton spectra with a freeze-out temperature $T_{fo}$ of 109~MeV,
and a transverse velocity $\beta_T$ of 0.77 describes the most central $\phi$ data
as well. A similar conclusion, with monotonically changing results for
$T_{fo}$ and $\beta_T$, holds for the less central events. At
higher transverse momenta, all particles lie above the blast wave
fits, which suggests that the dominant particle
production mechanism is no longer soft physics but is
giving way to the expected jet fragmentation.
A study of the ratio $R_{CP}$ of the $\phi$ provides a
critical new piece of information in understanding the anomalous
proton--to--pion ratio seen in central heavy ion collisions at RHIC,
since the $\phi$ is a meson with a mass similar to that of a
proton. The $R_{CP}$ value for the $\phi$ above a transverse momentum of
1.5~GeV/$c$ is about 0.6, similar to that of the pions
but inconsistent with the proton value of 1. This indicates that the
$\phi$ meson is being suppressed in this $p_T$ range 
for the more central collisions.
The lower \pt
blast wave fits imply hydrodynamic behavior at the hadronic stage, while
the inconsistency with simple hydrodynamics at higher $p_T$, is something one
would expect in the jet fragmentation region. This transition is an
important factor to consider as one begins to understand the mechanism
of particle production in central collisions at RHIC.
\clearpage

\begin{acknowledgments}
We thank the staff of the Collider-Accelerator and Physics
Departments at Brookhaven National Laboratory and the staff
of the other PHENIX participating institutions for their
vital contributions.  We acknowledge support from the
Department of Energy, Office of Science, Nuclear Physics
Division, the National Science Foundation, Abilene Christian
University Research Council, Research Foundation of SUNY, and
Dean of the College of Arts and Sciences, Vanderbilt
University (U.S.A), Ministry of Education, Culture, Sports,
Science, and Technology and the Japan Society for the
Promotion of Science (Japan), Conselho Nacional de
Desenvolvimento Cient\'{\i}fico e Tecnol{\'o}gico and Funda\c
c{\~a}o de Amparo {\`a} Pesquisa do Estado de S{\~a}o Paulo
(Brazil), Natural Science Foundation of China (People's
Republic of China), Centre National de la Recherche
Scientifique, Commissariat {\`a} l'{\'E}nergie Atomique,
Institut National de Physique Nucl{\'e}aire et de Physique
des Particules, and Institut National de Physique
Nucl{\'e}aire et de Physique des Particules, (France),
Bundesministerium f\"ur Bildung und Forschung, Deutscher
Akademischer Austausch Dienst, and Alexander von Humboldt
Stiftung (Germany), Hungarian National Science Fund, OTKA
(Hungary), Department of Atomic Energy and Department of
Science and Technology (India), Israel Science Foundation
(Israel), Korea Research Foundation and Center for High
Energy Physics (Korea), Russian Ministry of Industry, Science
and Tekhnologies, Russian Academy of Science, Russian
Ministry of Atomic Energy (Russia), VR and the Wallenberg
Foundation (Sweden), the U.S. Civilian Research and
Development Foundation for the Independent States of the
Former Soviet Union, the US-Hungarian NSF-OTKA-MTA, the
US-Israel Binational Science Foundation, and the 5th European
Union TMR Marie-Curie Programme.
\end{acknowledgments}

\begin{appendix}
\section{ Combinatorial Background}

Assume there are N tracks per event of which n are positive and N-n are negative, the
probability of the partition being $P(n)$ such that $\sum_{n=0}^{N} P(n) = 1$.

All expressions below refer to the average number of pairs per event
\vspace{0.5cm}
\subsection{ General relation between the number of like and unlike pairs}
The total number of pairs is:
\begin{equation}
       N_{pairs} = \frac{N(N-1)}{2}
\end{equation} 
The number of + - pairs is:
\begin{equation}
       n_{+-} = \sum_{n=0}^{N} n(N-n)P(n) = N \langle n  \rangle - \langle{n^2}\rangle
\end{equation} 
The number of + + pairs is:
\begin{equation}
       n_{++} = \sum_{n=0}^{N} \frac{n(n-1)}{2} P(n) =  
\frac{ \langle{n^2}\rangle - \langle{n}\rangle}{2}
\end{equation} 
The number of - - pairs is:
\begin{equation}
       n_{--} = \sum_{n=0}^{N} \frac{(N-n)(N-n-1)}{2} P(n)
\end{equation}
\begin{equation}
       n_{--} = \frac{N(N-1)}{2} - (N-\frac{1}{2})\langle{n}\rangle +
       \langle{n^2}\rangle
\end{equation} 
We therefore get:
\begin{equation}
     n_{++} +  n_{--} =  \frac{N(N-1)}{2} - (N \langle{n}\rangle -  \langle{n^2}\rangle)
\end{equation} 
The last expression is trivial: the number of like-sign pairs is equal to the total number pairs
minus the number of of unlike-sign pairs.\\
The ratio R of unlike to like pairs is therefore:
\begin{equation} 
    R = \frac{n_{+-}}{n_{++} +  n_{--}} = \frac{N \langle{n}\rangle - \langle{n^2}\rangle}{N(N-1)/2 - (N \langle{n}\rangle -  \langle{n^2}\rangle)}
\end{equation}  
This is a general result, free of any assumption and should always be fulfilled, in pure combinatorial 
background as well as in a mixture of signal + combinatorial background. 

\vspace{0.5cm}
\subsection{Combinatorial background}
Consider a pure combinatorial background sample. By its essence there are 
\underline{no correlations} between tracks in such a sample i.e. within an event 
the probability $p$ to have a positive (or a negative) track is  
constant and independent of the number of tracks. Therefore the probability $P(n)$ of having
$n$ positive tracks out of the total of $N$ tracks is given by a binomial distribution:
\begin{equation} 
    P(n) = \frac{N!}{n!(N-n)!} p^n (1-p)^{N-n}
\end{equation}
\begin{equation} 
   \langle{n}\rangle = pN
\end{equation}   
\begin{equation} 
   \langle{n^2}\rangle = \sigma^2 +  {\langle{n}\rangle}^2 = Np(1-p) + p^2N^2
\end{equation}    
Replacing these values in expression (6) gives:
\begin{equation}     
     R = \frac{(p-p^2)}{(1/2 - p + p^2)}
\end{equation}

  If there is charge symmetry, i.e. $p=0.5$ one gets $R=1$ and consequently the
combinatorial background is given by:
\begin{equation}     
     n_{+-}^{CB} = n_{++} +  n_{--} 
\end{equation}
   This is an exact relation. It holds with quite good accuracy 
even if there is some charge asymmetry. For example for an asymmetry of 10\% (i.e. $p=0.525) $R=0.995.
 
\vspace{0.5cm}
\subsection{The formula $N_{+-}^{CB} = 2 \sqrt{N_{++}N_{--}}$} 
The combinatorial background is rigorously given by this formula  
provided that the number N of tracks per event has a Poisson distribution:
\begin{equation}
   \mathcal{P}(N) = \frac{\langle{N}\rangle ^N e^{-N}}{N!}
\end{equation}

Again we assume that the N tracks are divided into n positive and (N-n) negative tracks,
the partition is given by the binomial distribution (A7) and
all expressions below refer to average number of pairs per event.\\
Number of + + pairs:
\begin{equation}
    N_{++} = \sum_{N=2}^{\infty} \mathcal{P}(N) \sum_{n=0}^{N} P(n) \frac{n(n-1)}{2}
\end{equation}       
Using relations (A7) and (A11) and some algebra leads to: 
\begin{equation}
     N_{++}   =  \frac{1}{2}p^2\langle{N}\rangle{^{2}}
\end{equation}
Similarly the number of - - pairs is given by:
\begin{equation}
    N_{--} = \sum_{N=2}^{\infty} \mathcal{P}(N) \sum_{n=0}^{N} P(n) \frac{(N-n)(N-n-1)}{2}
\end{equation}       
\begin{equation}
   \hspace{0.9cm}  =  \frac{1}{2}(1-p)^2\langle{N}\rangle{^2}
\end{equation} 
The number of combinatorial background + - pairs is:
\begin{equation}
     N_{+-}^{CB} =  \sum_{N=2}^{\infty} \mathcal{P}(N) \sum_{n=0}^{N} P(n) n(N-n)
\end{equation} 
\begin{equation}
   \hspace{0.9cm}  = p(1-p)\langle{N}^2\rangle
\end{equation} 
Inspecting (A15), (A16) and (A17) shows that: 
\begin{equation}
      N_{+-}^{CB} = 2 \sqrt{N_{++}N_{--}}
\end{equation}

\section{Data tables of centrality selected  $\phi$ spectra}

The invariant yields, $\frac{1}{2\pi m_{T}}~.~\frac{d^{2}N}{dm_{T}dy}$, of the
 $\phi$ mesons in different centrality bins are shown in Table~\ref{datatable}.

\begin{table*}
\caption{\label{datatable} $m_{T}$ spectra of $\phi$ mesons in different
centrality bins.
The systematic errors on invariant yields are from combinatorial background normalization, $\phi$ counting mass window,
acceptance correction efficiencies from Monte Carlo and occupancy dependent corrections.}
\begin{ruledtabular}
\begin{tabular}{cccccc}
Centrality&$m_{T}$&$m_{T}$ bin size&$\frac{1}{2\pi m_{T}} \frac{d^{2}N}{dm_{T} dy}$&Stat. error&Syst. error\\
(\%)&(GeV/$c^{2}$)&(GeV/$c^{2}$)&$[(GeV^{2}/c^{4})^{-2}]$&$[(GeV^{2}/c^{4})^{-2}]$& $[(GeV^{2}/c^{4})^{-2}]$\\\hline
0 - 10&1.365&0.4&0.51114&0.15120&0.0778747\\
&1.691&0.2&0.19114&0.04656&0.029121\\
&1.891&0.2&0.12065&0.02702&0.018382\\
&2.091&0.2&0.07313&0.01758&0.011142\\
&2.291&0.2&0.03335&0.01372&0.005081\\
&2.565&0.4&0.01683&0.00541&0.002564\\
&2.891&0.2&0.01353&0.00406&0.002061\\
&3.300&1.0&0.00268&0.00074&0.000408\\\hline
10 - 40&1.364&0.4&0.37232&0.04145&0.05667\\
&1.691&0.2&0.09224&0.01255&0.01404\\
&1.891&0.2&0.04872&0.00719&0.00742\\
&2.091&0.2&0.03867&0.00483&0.00589\\
&2.291&0.2&0.02320&0.00389&0.00353\\
&2.564&0.4&0.01246&0.00160&0.00189\\
&2.891&0.2&0.00415&0.00110&0.00063\\
&3.294&1.0&0.00120&0.00021&0.00018\\\hline
40 - 92&1.364&0.4&0.04432&0.00622&0.00645\\
&1.691&0.2&0.01295&0.00185&0.00189\\
&1.891&0.2&0.01009&0.00110&0.00147\\
&2.091&0.2&0.00531&0.00074&0.00077\\
&2.291&0.2&0.00287&0.00060&0.00042\\
&2.563&0.4&0.00126&0.00024&0.00018\\
&2.891&0.2&0.00062&0.00019&0.00009\\
&3.293&1.0&0.00019&0.00004&0.00003\\
\end{tabular}
\end{ruledtabular}
\end{table*}

\section{Systematic errors on $dN/dy$ and $T$}

The sources of the systematic errors on yield ($dN/dy$ and T) 
measurements are as follows:

a) Systematic error on the combinatorial background normalization, $\delta_{norm}$:
This originates from the systematics of the event mixing. Since both
same event and mixed event like sign distributions represent pure
combinatorials, we estimated unlike sign combinatorial background by normalizing the mixed event unlike sign distributions
to 2$\sqrt{N_{++}^{Same~event}\cdot N_{--}^{Same~event}}$ and 2$\sqrt{N_{++}^{Mixed~event}\cdot N_{--}^{Mixed~event}}$ and
the difference in the extracted $\phi$ signal from the real data
between these two normalizations are attributed as the systematic uncertainty.

We use the same normalization factor for all $m_{T}$ bins. So, the above
systematics are applicable to the measured $dN/dy$ only, not on the
inverse slope, $T$.

b) Systematics of $\phi$ mass window, $\delta_{mass}$: We count the number of reconstructed
$\phi$ mesons by integrating the $\phi$ meson invariant mass
spectra within $\pm$ 5 MeV mass window  with respect to the measured centroids
in both data and Monte Carlo. The systematic associated with
this mass window is estimated by measuring the extent of the changes in $dN/dy$ and $T$ after
constructing $\phi$ meson $m_{T}$ spectra within five different mass windows
$\pm$ 3, $\pm$ 5, $\pm$ 8, $\pm$ 10 and $\pm$ 15 MeV, with respect to the
measured $\phi$ centroids.

c) Uncertainties in extrapolation of
$\phi$ meson $m_{T}$ spectra to $m_{T}$ = $m_{\phi}$ , $\delta_{extrap}$: This is studied by \\
(i) fitting the $m_{T}$ distributions
using exponential and Boltzmann functions. two different fitting functions,
exponential and Boltzmann, and \\
(ii) fitting the transverse mass spectra within different $m_{T}$ ranges.
These are applied to both $dN/dy$ and $T$.

d) Acceptance correction systematics, $\delta_{MC}$:
The systematics associated with acceptance correction factors 
derived from Monte Carlo analysis are investigated by considering two sources:\\
(i) Tuning of detector alignments in Monte Carlo with reference to
the real data ($\sim$ 3\%), and\\
(ii) Systematics in the fiducial geometries in data and Monte Carlo ($\sim$ 12\%).

The systematic error from this source is independent of  the momenta of
reconstructed $\phi$ mesons. So, this is attributed to $dN/dy$ only.

e) Systematic error in the occupancy dependent efficiency corrections, $\delta_{occu}$:
The systematic error associated with this efficiency is estimated by
calculating the occupancy dependent correction with different track confirmation criteria and
is independent of the pair momenta. This is a systematic effect on 
$dN/dy$ only.

The above systematic errors are quoted in Tables~\ref{systematics1} and
\ref{systematics2} for $dN/dy$ and $T$, respectively.

\begin{table}
\caption{\label{systematics1} Systematic error in $dN/dy$.}
\begin{ruledtabular}
\begin{tabular}{ccccccc}
Centrality&$\delta_{norm}$&$\delta_{mass}$&$\delta_{extrap}$&$\delta_{MC}$&$\delta_{occu}$&$\delta_{sys}^{tot}$\\
(\%)&(\%)&(\%)&(\%)&(\%)&(\%)&(\%)\\\hline
Minimum Bias     &0.7&2.6&4.0&12.4&7.8&15\\
0 - 10 &0.8&4.2&5.2&12.4&10&17\\
10 - 40&1.1&2.3&5.9&12.4&8.5&16\\
40 - 92&0.7&3.1&6.0&12.4&7&16\\
\end{tabular}
\end{ruledtabular}
\end{table}

\begin{table}
\caption{\label{systematics2} Systematic error in $T$.}
\begin{ruledtabular}
\begin{tabular}{cccc}
Centrality&$\delta_{mass}$&$\delta_{fit}$&$\delta_{sys}^{tot}$\\
(\%)&(\%)&(\%)&(\%)\\\hline
Minimum Bias    &0.6&4.9&5\\
0 - 10 &1.1&5.2&5\\
10 - 40&1.1&6.2&6\\
40 - 92&1.1&4.2&4\\
\end{tabular}
\end{ruledtabular}
\end{table}

\end{appendix}
\clearpage


\end{document}